\documentclass[preprint]{aastex}

\newcommand{\Ha}{H$\alpha$}
\newcommand{\Hb}{H$\beta$}
\newcommand{\Hg}{H$\gamma$}
\newcommand{\feii}{\ion{Fe}{2}}
\newcommand{\Few}{\ion{Fe}{2}\,$\lambda$4570}
\newcommand{\FeIIHb}{\ion{Fe}{2}\,$\lambda$4570/H$\beta$}
\newcommand{\OII}{[\ion{O}{2}]\,$\lambda$3727}
\newcommand{\OIIItwo}{[\ion{O}{3}]\,$\lambda\lambda$4959,5007}
\newcommand{\OIIIone}{[\ion{O}{3}]$\,\lambda$4959}
\newcommand{\OIII}{[\ion{O}{3}]$\,\lambda$5007}

\newcommand{\NII}{[\ion{N}{2}]}

\newcommand{\MgII}{\ion{Mg}{2}\,$\lambda\lambda$2795,2802}

\newcommand{\Lx}{L$_{\rm X}$}
\newcommand{\LFIR}{L$_{\rm FIR}$}
\newcommand{\kms}{\rm km\,s$^{-1}$}
\newcommand{\calP}{{\cal P}}
\begin{document}

\title{ Spectroscopic Properties of QSOs Selected from \\
        Ultraluminous Infrared Galaxy Samples}

\author{X.Z. Zheng}
  \affil{National Astronomical Observatories, Chinese Academy of Sciences, \\
         Datun Road A20, Chaoyang District, Beijing 100012, China}
  \email{zxz@alpha.bao.ac.cn}
\author{X.Y. Xia}
  \affil{Department of Physics, Tianjin Normal University, \\
         Weijin Road, Tianjin 300074, China}
  \email{xyxia@bac.pku.edu.cn}
\author{S. Mao}
 \affil{Univ. of Manchester, Jodrell Bank Observatory,
          Macclesfield, Cheshire SK11 9DL, UK}
 \email{smao@jb.man.ac.uk }
\author{H. Wu}
  \affil{National Astronomical Observatories, Chinese Academy of Sciences, \\
        Datun Road A20, Chaoyang District, Beijing 100012, China}
  \email{wu@bac.pku.edu.cn}
\and
\author{Z.G. Deng}
  \affil{Department of Physics, Graduate School, Chinese Academy of Sciences,
	Beijing 100080, China}
  \email{dzg@bac.pku.edu.cn}

\begin{abstract}

We performed spectroscopic observations for a large infrared
QSO sample with a total of 25 objects. The sample was compiled from
the  QDOT redshift survey, the 1~Jy ULIRGs survey and  a sample
obtained by a cross-correlation study of the IRAS Point Source Catalogue with
the ROSAT All Sky Survey Catalogue.  Statistical analyses of the optical
spectra show that the vast majority of infrared QSOs have narrow
permitted emission lines (with FWHM of \Hb\ less than 4000~\kms) and more
than 60\% of them are luminous narrow line Seyfert 1 galaxies.  Two of
the infrared QSOs are also classified as low ionization BAL QSOs.
More than 70\% of infrared QSOs are moderately or extremely strong \feii\
emitters.
This is the highest percentage of strong \feii\ emitters in
all subclasses of QSO/Seyfert 1 samples. 
We found that the \feii\, to \Hb\, line ratio
is significantly correlated with the \OIII\ peak
and \Hb\ blueshift. Soft X-ray weak infrared QSOs tend to
have large blueshifts in permitted emission lines and significant
\feii\,48,\,49 (5100--5400~\AA) residuals relative to the Boroson \&
Green \feii\ template. If the blueshifts in permitted lines are
caused by outflows, then they appear to be common in infrared QSOs.
As the infrared-selected QSO
sample includes both luminous narrow line Seyfert 1 galaxies and
low ionization BAL QSOs, it could be a useful laboratory to investigate
the evolutionary connection among these objects.

\end{abstract}

\keywords{galaxies: Seyfert --- quasars: emission lines --- quasars: general}

\section{Introduction}

Two of the most important reasons for investigating the ultraluminous
IRAS galaxies (ULIRGs) are to find the evolutionary connection between
circumnuclear massive starbursts and active galactic nuclei (AGNs)
and to identify the evolution path from galaxy mergers to elliptical
galaxies and QSOs (see Sanders \& Mirabel 1996 for a review).
In recent years, significant progress has been achieved with both
space and ground-based telescopes (Surace et al. 1998;
Surace, Sanders, \& Evans 2000; Genzel et al. 1998; Farrah et al. 2001
and reference therein).  It is now widely accepted that the vast
majority ($\ga 95\%$) ULIRGs are strongly interacting and merging
galaxies while some of them are post-merger galaxies. The AGN
phenomenon probably appears at the final merging stage
(e.g., Clements et al. 1996;  Kim, Veilleux, \& Sanders 1998;
Zheng et al. 1999; Canalizo \& Stockton 2001a; Cui et al. 2001).
Spectral analyses for large samples of ULIRGs reveal that the fraction
of objects with AGN spectral characteristics
is about 25--30\% while the fraction of
QSOs/Seyfert 1s is less than 10\% (Wu et al. 1998; Lawrence et al. 1999).
However, the percentage of QSOs/Seyfert 1s increases with increasing
infrared luminosity, reaching 30--50\% for $L_{\rm IR} > 10^{12.3}L_{\odot}$
(Veilleux, Kim, \& Sanders 1999).  As the infrared luminosity of ULIRGs
is equivalent to the bolometric luminosity of optically selected QSOs
(Sanders 2001),  we find it convenient to refer to
QSOs/Seyfert 1s selected from ULIRGs as IR QSOs throughout this paper.

Previous spectroscopic studies of small and statistically incomplete
IR QSO samples have uncovered some unusual properties compared with
optically selected QSOs/Seyfert 1s.  Many IR QSOs are extremely strong
\feii\ emitters (\FeIIHb\ $ > 2.0$), for example, PHL~1092, IRAS~07598+6508
and Mrk 231 (L\'{\i}pari  1994; Lawrence et al. 1997).
In fact, almost 100\% of extremely strong \feii\ emitters are luminous
IR QSOs (L\'{\i}pari et al. 2002). More than twenty years after their
discovery, the origin of such extremely strong optical \feii\ emissions
in QSOs/Seyfert 1s is still being debated.  It has become clear, however,
that the strength of the \feii\ emission cannot be explained in the
framework of photoionization excitation. If strong outflows and
shocks are present in strong/extremely strong \feii\ emitters,
non-radiative shock heating and overabundance of iron may help
to explain the strong \feii\ emission (Collin \& Joly 2000).
Studies of IR QSOs may therefore shed new light on the origin
of the \feii\ emission.

Furthermore, the fraction of low ionization broad absorption line QSOs
(lo-BAL QSOs) is much higher in an IR QSO sample (27\%) than that in an
optically selected QSO sample (1.4\%) (Boroson \& Meyers 1992).
The lo-BAL QSOs are defined as a subclass of broad absorption line QSOs
with an obvious \MgII\ doublet and low-ionization line
absorption troughs (Weymann et al. 1991).
Some IR QSOs belong to
yet another class of unusual AGNs, the luminous narrow line Seyfert 1s
(hereafter NLS1s, Osterbrock \& Pogge 1985).
NLS1s are defined by their optical emission line properties.
They have narrow hydrogen Balmer lines with typical full width at half
maximum (FWHM) $\approx$ 500--2000~\kms. The \OIII/\Hb\ line ratio is
less than 3 and most of them have strong  \feii\ emissions.
In the X-ray band, NLS1s show systematically steeper power-law slopes
in the continuum than normal Seyfert 1s. Some NLS1s exhibit rapid
X-ray variabilities as well (Boller, Brandt \& Fink 1996). Moran, Halpern, \& Helfand (1996)
pointed out that many NLS1s are luminous in the infrared band.
IR QSOs therefore offer a unique opportunity to
investigate potential physical connections between IR QSOs,
lo-BAL QSOs and luminous NLS1s (Brandt \& Gallagher 2000; Canalizo \&
Stockton 2001b; Sanders \& Mirabel 1996). 

In this paper, we study the spectroscopic properties of IR QSOs based
on a large sample of 25 objects. We compare their properties with
those of the  Boroson \& Green (1992, hereafter BG92) sample, which 
includes 87 optically selected QSOs.
The outline of the paper is as follows. In \S 2,
we discuss how our IR QSO sample is compiled.  The observations and data
reduction are described  in \S 3.  We present the spectra and
statistical properties of our sample IR QSOs in \S 4 and finally,
in \S 5, we discuss and summarize our results.  Throughout
this paper we use a Hubble constant of $H_0$\,=\,50~\kms\,Mpc$^{-1}$ and
$\Omega_0$\,=\,1 and no cosmological constant. As all our objects
have redshift lower than 0.35, the adoption of a different density
parameter and cosmological constant has little effects on our results.

\section{Sample Selection}

Our IR QSO sample is compiled mainly from
three sources
\begin{itemize}
\item
The QDOT redshift survey is a survey of the 
IRAS galaxies sparse-sampled at a
rate of one in six.  It includes 2387 IRAS galaxies complete down to
a 60~$\mu$m flux density limit of 0.6 Jy. Lawrence et al. (1999) gave
a table of 97 ULIRGs (for H$_0$\,=\,50~\kms\,Mpc$^{-1}$ and $\Omega_0$\,=\,1)
with the criterion that the 60\,$\mu$m  luminosity is greater
than $10^{12}$\,L$_\odot$.  The optical spectral features and
classifications are given in the ULIRG table. There are eight objects
in the table identified as IR QSOs.
\item
Kim \& Sanders (1998) selected 118 ULIRGs (infrared luminosities
L(8--1000~$\mu$m)\,$>$\,10$^{12}$\,L$_{\odot}$ for
H$_0$\,=\,75~\kms\,Mpc$^{-1}$
and $\Omega_0$\,=\,0) from the 
criterion of 60~$\mu$m flux density greater than 1 Jy in sky region
of $\delta > -40\degr$ and $\left|b\right|\,>\,30\degr$. Veilleux et al.
(1999) gave the spectroscopic features and classification of 108 out
of these 118 ULIRGs. There are 10 IR QSOs among these 108 ULIRGs.
\item
Moran et al. (1996) presented spectroscopic classifications
for a catalogue of IRAS galaxies selected from the cross-correlation of
the IRAS Point Source Catalogue with the ROSAT All Sky Survey by Boller
et al. (1992).  This catalogue consists of 241 objects and 80 of them
are identified as QSOs/Seyfert 1s. 11 of these objects are IR QSOs.
\end{itemize}

Due to the constraint of the observatory site and instrumental capability,
we selected our targets by requiring the IR QSOs to be in the northern
sky ($\delta > -30\degr$) and z$ < $0.35 with
L(8--1000~$\mu$m)\,$>$\,10$^{12}$\,L$_{\odot}$
for H$_0$\,=\,50~\kms\,Mpc$^{-1}$ and $\Omega_0$\,=\,1.  In addition,
we included F09427+1929 (Zheng et al. 1999).  Taking into account the
overlapping sources, the sample consists of 25 IR QSOs.
Table~\ref{tab1} lists the basic parameters.
Note that all the infrared luminosities have been converted using
H$_0$\,=\,50~\kms\,Mpc$^{-1}$ and $\Omega_0$\,=\,1.

While the sample size is still moderate, it is interesting
to put this number in the context of the total expected number of IR QSOs
in the local universe. The PSCz catalogue
provides a complete redshift survey of 15411 IRAS galaxies
(Saunders et al. 2000). About 900 ULIRGs were found, which implies
the percentage of ULIRGs in the IRAS galaxy catalogue is about 6\%.
The percentage of IR QSOs among ULIRGs is approximately 10\%, as 
there are 8 IR QSOs among 97 ULIRGs in the QDOT
catalogue and 10 IR QSOs among 108 ULIRGs in the 1 Jy
sample. So the fraction of IR QSOs in the complete PSCz catalogue
is of the order of 0.6\%, i.e., the total IR QSOs in the
PSCz catalog may be less than 100. Our sample therefore includes
roughly one quarter of IR QSOs in the local universe.
Statistical results based on this quite large IR QSO
sample should be representative.

\section{Observations and Data Reduction}

Long-slit optical spectroscopic observations were carried out on the
2.16m telescope at the Xinglong station of the National Astronomical
Observatories. The observations were mostly performed between October
1998 and November 1999 using an OMR spectrograph while some preliminary
studies were conducted before 1998. For our 1998 and 1999 observations,
a Tektronix 1024$\times$1024 CCD was used giving a wavelength coverage
of 4000~\AA\ to 9000~\AA\ with a grating of 200~\AA\,mm$^{-1}$. The
spectral resolution was 9.7~\AA\ (2 pixels). The slit width varied from
1.5\arcsec\ to 3.5\arcsec\ to match the seeing at the Xinglong station.
For the pre-1998 observations, the instrument setup was slightly
different. For these observations, a grating of 195~\AA\,mm$^{-1}$
was used and the coverage is from 3500~\AA\ to 8100~\AA\ with a
resolution of 9.3~\AA.  For F01572+0009, the spectrum coverage was in
the range of 3800\,$\sim$\,6300~\AA\ with a resolution of 4.9~\AA.
The observation log is given in Table~\ref{tab2} listing observation
epochs, exposure times, approximate seeings and adopted slit widths.

Data reduction was performed using IRAF software. CCD
reductions included bias subtraction, flat field correction
and cosmic-ray removal. Sky light subtractions were accomplished
during the extraction procedure. Wavelength calibrations were
carried out using a He-Ar lamp. The resulting wavelength accuracy
is better than 1~\AA. KPNO standard stars were observed
for flux calibrations. The telluric O$_2$ absorption bands
near 6870~\AA\ and 7620~\AA\ were removed using the spectra of
the standard stars.

The IRAF package `SPLOT' is used to measure isolated emission lines.
`SPECFIT'\footnote{SPECFIT is developed and kindly provided by Gerard A.
Kriss.}, an interactive spectral analysis procedure linked to IRAF is
used to measure blended lines  (e.g., \Ha\
and \NII$\lambda\lambda$6548,6583).
It can match a wide variety of emission lines, absorption lines,
and continuum models. We model the emission lines with Gaussian
profiles and the local continuum as a power-law.

Since the intrinsic extinction of ULIRGs is significant, the Galactic
extinction is ignored in our analysis. The extinction
correction is calculated approximately
according to Veilleux \& Osterbrock  (1987) and the
intrinsic broad line \Ha/\Hb\ ratio is taken to be 3.1 (Baker 1997). 
The \Hb/H$\gamma$ ratio is used to estimate the extinction when \Ha\ is 
out of the spectral coverage (Osterbrock 1989).
The measured fluxes of Balmer lines are the sum of the narrow and broad components.
It is noted that the intrinsic broad Balmer line ratios may be larger than the
adopted values (e.g. MacAlpine 1985) and the extinction
corrections may have large uncertainties. Fortunately, the emission line 
ratios adopted in our
analyses, e.g. \FeIIHb\ and \OIII/\Hb, are almost independent of
extinction, due to the adjacency of the involved lines.
For the \feii\ residual measurement, the uncertainty from
the extinction correction will be discussed in section 4.3.
In general, the uncertainty for the
flux measurement introduced by the extinction correction is less than 20\%.

As the \feii\ emission is moderately or extremely strong
for most of our sample galaxies, \feii\ multiplets seriously blend
with the \Hb\ and \OIIItwo\ lines and contaminate the continuum.
We carefully remove the \feii\ multiplets following BG92. Their method
uses an \feii\ template derived empirically from high quality data of
I ZW 1, a typical NLS1 galaxy. We then measure the flux for all
emission lines based on the \feii-subtracted continuum. On the other hand,
the emission lines for a quite large fraction of our
sample IR QSOs show a remarkable
asymmetric profile. In such cases, double-Gaussian profiles are needed
to fit them. The asymmetry and blueshift can then be measured
from the fitting. In the following subsections
we describe in more detail the \feii\ multiplets removal
method and the measurement of the line asymmetry and blueshift.

\subsection{\feii\ Multiplets Removal}

In order to estimate the \feii\ strength and measure the line fluxes reliably,
we adopt the BG92 method which relies on an \feii\ template. The template and
observed spectra are both transformed into the rest frame. The
template is broadened by convolving with a Gaussian of various
line widths and scaled by multiplying by a factor indicating the
line strength.
The best match is then searched for in the two-dimensional parameter
space of the line width and line strength. 
A good \feii\ subtraction is found when the parts of the continuum
between the  \Hg\ and \Hb\ and between
5100--5400~\AA (which covers the \feii\ multiplets 48,\,49)
are flat.
The best-fit \feii\ template emission lines are then subtracted from the  
observed spectrum.  The \feii\ flux is then
determined from the best fitting  \feii\
template between the rest wavelengths 4434~\AA\ and 4685~\AA.
The procedure of \feii\ subtraction is illustrated in Fig.~\ref{figspectra}.
In each panel of Fig.~\ref{figspectra}, 
the top curve is the dereddened spectrum,
the bottom curve is the \feii\ template while the middle curve shows
the \feii-subtracted spectrum. Note that the \feii-subtracted spectra
have been shifted downwards for clarity.

In the BG92 method, it is assumed that the relative strengths of the
\feii\ lines (within each multiplet and among multiplets) are the same
for different objects. For half of our spectra, the \feii\ emission
can be subtracted very well by the \feii\ template. However, for
F00275$-$2859, IRAS~07598+6508, F09427+1929, Mrk~231 and F20036$-$1547 ,
the \feii\ multiplets 48,\,49 (5100--5400~\AA) are stronger than
\feii\ multiplets 37,\,38 relative to the BG92 template.  In contrast,
for F02065+4705, F10026+4347, Z11598$-$0112, F20520$-$2329 and
F22454$-$1744, the \feii\ multiplets 37,\,38 (4500--4680~\AA) are
stronger than \feii\ multiplets 48,\,49 relative to BG template.
For these objects, significant \feii\ residuals can be seen in their
\feii-subtracted spectra. The extreme case is Z11598$-$0112,
for which nearly half of  the \Few\ (i.e. \feii\ multiplets 37,\,38)
is left after the multiplets 48,\,49 are subtracted. 
Fig.~\ref{FeIIresiduals} shows
two examples of such remarkable \feii\ emission residuals in
the \feii\ multiplets 48,\,49 and 37,\,38
in the top and bottom panels, respectively. 
Compared with the optical QSO sample of BG92,
the IR QSO sample contains more objects showing large deviations from
the \feii\ template.

The spectra For F13218+0552 and F23411+0228 are too noisy to detect the
\feii\ lines reliably, so we take the \feii\ strength for
F13218+0552 from Remillard et al. (1993) and ignore F23411+0228 in our
statistical analysis concerning the \feii\ strength.

\subsection{Emission Line Fitting and Measurement}

As a first step, we use a single Gaussian profile to fit each emission
line for all target galaxies. It works well for most emission lines.
However, a single Gaussian profile cannot fit some lines, e.g.,
those with
asymmetric profiles. In such cases, two Gaussian components are used
to fit the emission lines, namely, one narrow component and one
centroid-shifted broad component. The blueshift (or redshift) is
defined as the shift of the broad component relative to the narrow component
in units of \kms.  Fig.~\ref{figtemplate} illustrates the two Gaussian
component fitting for the permitted line \Hb\ (top panel)
and forbidden line \OIII\ (bottom panel).
In Fig.~\ref{figtemplate}, the observed  and fitted profiles
are shown by the solid and dashed lines, respectively, while the
dot-dashed line is for the Gaussian components and the dotted line
represents the fitting residual.  As one can see, for F01572+0009,
the blueshift of the permitted emission line \Hb\ of F00275$-$2859
is 750~\kms\ while the blueshift of \OIII\ is 510~\kms.

Moreover, there is no correlation between the FWHM of the
narrow Gaussian component
of the permitted line \Hb\ and that of the forbidden line \OIII.
It implies that the narrow Gaussian component of the permitted emission lines
is different from the narrow lines (such as \OIII)
from the narrow line region. Efforts were
also made to separate the narrow Gaussian component contributed by the
narrow line region. However, the contribution from the narrow line region
is usually less than 3\%  so we ignore this component in our discussions.

We also measure the asymmetry parameter defined by de Robertis (1985):
\begin{equation}
{\rm asy} = {\lambda_{\rm c}(3/4) - \lambda_{\rm c}(1/4) \over
                     \Delta\lambda(1/2)},
\end{equation}
where $\lambda_{\rm c}(1/4)$ and $\lambda_{\rm c}(3/4)$ are the wavelength
centers at  $1/4$ and $3/4 $ of the maximum, respectively, and
$\Delta\lambda(1/2)$ is the FWHM.  The asymmetry parameter is positive
(negative) if there is excess light in the blue (red) wing.

Based on the \feii-subtracted spectra and Gaussian fitting of emission lines,
we measured the flux, FWHM,
and equivalent width (EW) for each strong emission line for all our objects.
The fluxes of broad lines or asymmetric lines refer to the sum of double 
components.
For most targets, the uncertainty of flux measurement of emission
lines is about 10\%, but for low S/N cases it could be up to 20\%.
The uncertainty of the blueshift measurement is within 150 \kms.
However, for three sources (F02054+2835, F13218 +0552 and F23411+0228),
the fluxes of emission lines have large uncertainties due to
the poor S/N and hence they are less reliable.

The dereddened and \feii-subtracted spectrum for each sample IR QSO
is shown in Fig.~\ref{figspectra}.  Table~\ref{tab3} lists the FWHM,
blueshifts and asymmetry parameters for \Hb\ and \OIII.
The intrinsic FWHM values are obtained from
the subtraction, in quadrature, of the observed FWHM and that of the
instrumental profile, measured from the comparison lamp lines. Note that
only the significant \OIII\ blueshifts (the \OIII\ blueshift $>$ 500~\kms)
are listed.  In Table~\ref{tab4}, color excess E(B$-$V), the equivalent
widths of the emission lines and various line ratios are listed.

\section{Statistical Results}

As discussed in the introduction, our sample is a unique one
to investigate the physical connection
among IR QSO, luminous NLS1s and lo-BAL QSOs.  We therefore
performed statistical studies similar to those in BG92 for their optical
sample, which includes 87 QSOs from the Bright Quasar Survey Catalogue
with redshift less than 0.5 (Schmidt \& Green 1983). The results
for these two samples will be compared in order to
understand any
possible evolutionary connection between IR QSOs and classical QSOs.

\subsection{The Percentage of NLS1s and Strong \feii\ Emitters}

It is obvious from Table~\ref{tab3} that for all our IR QSOs
except F16136+6550 and F18216+6419, the \Hb\ FWHM is less than 4000~\kms.
This differentiates IR QSOs from
classical QSOs as the main characteristic of
classical QSOs is the presence of broad permitted emission lines
with typical FWHM between 4000 and 10000~\kms\ (Rodr\'{\i}guez-Ardila
et al. 2000). To make it clear, Fig.~\ref{fighistogram1} shows the
distribution of the \Hb\ FWHM for our sample (top) and the BG92 sample
(bottom). A Kolmogorov-Smirnov (K-S) test indicates a probability of
$9.5  \times 10^{-5}$ for the two distributions being the same.
 From Table~\ref{tab3} and Fig.~\ref{fighistogram1}, the
percentage of IR QSOs with \Hb\ FWHM less than 2000~\kms\ in our
sample is 60\% (15/25).
In comparison, only 23\%  of BG92 QSOs have FWHM \Hb\ less than 2000~\kms\
and one third of BG92 QSOs are classical QSOs with \Hb\  FWHM
larger than 4000~\kms. 

As part of our IR QSOs are selected from the IRAS-ROSAT cross-correlation
catalogue, our sample may be biased to include more NLS1s (Stephens 1989).
To check this, we performed statistics for the sub-sample
of 15 objects selected from two purely IR-selected samples
(the QDOT redshift survey and 1 Jy ULIRG sample).
Out of these 15 objects, 8 are identified as NLS1s (53\%). Therefore, there is
no clear difference in the fraction of NLS1s between the whole sample and the
purely IR-selected sub-sample.

We also carefully investigated whether the \Hb\ blueshift could influence
the \Hb\ FWHM and concluded that this possibility is unlikely.
Therefore the percentage of NLS1s seems genuinely high in our IR QSO sample.

Fig.~\ref{fighistogram2} shows the histograms of the \FeIIHb\ ratio
for the IR QSO and
BG92 samples. We can see from Fig.~\ref{fighistogram2} that the \FeIIHb\
ratio distributions for the two samples are quite different. A K-S test
reveals that the
probability for these two distributions being the same is $3.0 \times 10^{-6}$.
Table~\ref{tab4} shows that 25\% (6 in 24) and 46\% (11 in 24) of IR QSOs are 
respectively extremely
strong (\FeIIHb\ $ > 2.0$) and moderately strong ($1.0 <$ \FeIIHb\ $ < 2.0$)
\feii\ emitters (Joly 1991; V\'{e}ron-Cetty, V\'{e}ron-Cetty,
\& Con\c{c}alves 2001).  In contrast, for the BG92 sample, only 15 out
of 87 (17\%) QSOs are moderately strong \feii\ emitters and there are no
extremely strong \feii\ emitters present.  The high percentage of
strong \feii\ emitters in our sample can also be seen by comparing with
that of the overall AGN population.  V\'{e}ron-Cetty et al. (2001)
found that moderately strong \feii\ emission occurs in only
about 5\% of AGNs and only a few AGNs are extremely
strong \feii\ emitters.  Furthermore, the two weakest \feii\ emitters
(with \FeIIHb\ $ < 0.5$) in our sample (F16136+6550 and
F18216+6419) have  broad permitted emission lines with FWHM larger
than 4000~\kms, and hence are classical QSOs.

\subsection{The Correlations}

In this subsection, we study the correlations between various emission
lines and continua for our sample IR QSOs. For this purpose, we
performed Spearman Rank-order (S-R) correlation analyses among various
quantities and investigate the implications of these correlations.
As mentioned above, most of our sample IR QSOs are moderately/or
extremely strong \feii\ emitters, and the origin of the \feii\
emission is still not understood. Hence
our analysis focuses on the correlations of the \feii\
strength with other parameters.
Throughout this paper, the correlations are characterized by the
probability $\calP$ that the null hypothesis of no correlation is true.

Fig.~\ref{figewoiii_feiihb} shows the \FeIIHb\ line ratio
versus the \OIII\ peak, defined
as the peak height of the \OIII\ line relative 
to that of \Hb\ as in PG92. These two
parameters are anti-correlated with a correlation coefficient of 0.71
at a very high significance level ($\calP = 6.1 \times 10^{-6}$). This
result is consistent with BG92 although the correlation between these two
parameters is stronger than the one reported by BG92 
for optically selected QSOs.
The \FeIIHb\ ratio is also well correlated with the \OIII/\Hb\ ratio with
$\calP = 9.8 \times 10^{-3}$. Note that 6 IR QSOs in our sample have very weak
 \OIII\ emission (with EW of \OIII\ less than 5~\AA). Such a
weak \OIII\ emission is an
important characteristic of lo-BAL QSOs (Boroson \& Meyers 1992);
we return to this point in \S 5.2.

Fig.~\ref{figbs} shows the striking correlation between the \FeIIHb\
line ratio and the \Hb\
blueshift.  The S-R correlation coefficient is 0.75
which is highly significant with $\calP = 2.7 \times 10^{-5}$.
In particular, it can be seen from Table~\ref{tab3} and Fig.~\ref{figbs}
that about two thirds of our targets show \Hb\ blueshifts with
values as large as 2000~\kms.

As there are striking correlations between both the \FeIIHb\ and
\OIII\ peak and \FeIIHb\ and \Hb\ blueshift, we also performed a
correlation analysis between the \OIII\ peak and \Hb\ blueshift.
We found that these two parameters are well correlated
with an S-R correlation coefficient of 0.72 corresponding to
$\calP = 1.4 \times 10^{-4}$.  These tight correlations
between the \Hb\ blueshift, \FeIIHb\ and the \OIII\ peak
must reflect some physical connection between these lines.

Fig.~\ref{figfwhm_feiihb} plots  the \FeIIHb\ line
ratio versus the \Hb\ FWHM for 24 of our
objects (F23411+0228 is excluded due to its low S/N).
If the two classical QSOs with  \Hb\ FWHM larger than 4000~\kms are
excluded, then the S-R correlation coefficient between
these two parameters is 0.47 with
 $\calP =2.8 \times 10^{-2}$.  If we further exclude the 3 objects
with \Hb\ FWHM larger than 3000~\kms (but smaller than
4000~\kms), the correlation becomes
somewhat stronger, with an
S-R correlation coefficient of 0.58 corresponding to
$\calP = 9.9 \times 10^{-3}$.  We caution, however,
that if we include the 2 objects with \Hb\ FWHM larger than 4000~\kms,
the \FeIIHb\  line ratio and the \Hb\ FWHM are no longer well correlated.

Fig.~\ref{figasyfeii} shows the \FeIIHb\ line ratio versus the \Hb\ asymmetry parameter
defined by de Robertis (1985, see also eq. 1).  The S-R correlation
coefficient is 0.46 with $\calP=2.8 \times 10^{-2}$.
Our result is broadly consistent with BG92, although the \Hb\ asymmetry
parameters for our sample IR QSOs seems to be much larger than those for
the BG92 sample.  Quantitatively, just 13\% of BG92 QSOs have the
asymmetry parameter larger than 0.1, while in our sample 48\% (26\%) of
IR QSOs have the \Hb\ asymmetry value larger than 0.1 (0.2).

As described in \S 3.2, the \Hb\ blueshift value is determined by the
blueshift of the broad Gaussian component relative to the narrow Gaussian
component in permitted emission lines. The blueshifted broad Gaussian
component may be connected with the outflow of clouds in the
broad line region (Leighly 2001). Outflows
with a velocity of several hundred or even a few thousand~\kms\ could
produce shocks that can excite the emission lines. Such shocks
may be an important ingredient for understanding the strong correlation
seen between the \FeIIHb\ ratio and the \Hb\ blueshift.

\subsection{The Soft X-ray Properties}

We collected all the X-ray information from Moran et al. (1996) and
Xia et al. (2001) based on ROSAT archive data. Table~\ref{tab1}
lists the ratio of the soft X-ray luminosity to the infrared luminosity,
\Lx/\LFIR. This is an important quantity as
nearly all the luminosity of an ULIRG is emitted in the far-infrared
band (Surace et al. 2000).
We can see from Table~\ref{tab1}
that 17 of our IR QSOs were detected by either
ROSAT All Sky Survey (13/17) or pointing observations
(4/17). The ratio of the soft X-ray luminosity to the far-infrared
luminosity spans about four orders of magnitude.
Not surprisingly, the four objects
(IRAS~07598+6508, Mrk~231, F00275$-$2859 and F21219$-$1757)
detected only by ROSAT pointings have the lowest \Lx/\LFIR$<0.01$ values
compared with the other 14 objects detected by the ROSAT All Sky Survey.
For the non-detections, we calculated their upper 
limits of the soft X-ray luminosity as follows.
We assume a detection limit is 6 source photons (as in
the ROSAT All-Sky Survey Faint Source Catalogue, Voges et al. 2000) and an
average exposure time of 400 seconds for each source. The spectrum
of each source is taken to be a power-law with a photon index of 2.3
(appropriate for AGNs) and 
Galactic neutral hydrogen column density is adopted.
The estimated \Lx/\LFIR\ upper limits are
also given in Table 1. As might be expected,
 they all satisfy \Lx/\LFIR$<0.01$.

Fig.~\ref{figxraybs} shows the \Hb\ blueshift versus the ratio of
soft X-ray luminosity to far-infrared luminosity for 14 objects 
detected by the 
ROSAT All Sky Survey and ROSAT pointings.  In this statistic and
also in the following statistics concerning \Lx/\LFIR, we
excluded F16136+6550, F18216+6419 because they have broad
emission lines and are QSOs; F23411+0228 is also
excluded due to its low S/N spectrum. It is clear from
Fig.~\ref{figxraybs} that the
objects with larger \Hb\ blueshifts tend to be soft X-ray weak
and vice versa (the S-R correlation coefficient --0.61 
corresponding to $\calP =2.0 \times 10^{-2}$).  In 
Fig.~\ref{figxraybs}, we also labelled the
two potential low-BAL QSOs (F09427+1929 and F20036$-$1547, see
section 5.2) using their 
soft X-ray luminosity upper limits. These two objects
strengthen the trend described above.

During our data reduction, we noticed that the \feii\ emission
cannot be  subtracted very well by the \feii\ template of
BG92 for one third of our
targets, i.e., there are large residuals of \feii\ multiplets
48,\,49 (5100--5400~\AA) or \feii\ multiplets 37,\,38 (4500 --
4680~\AA) in the \feii-subtracted spectra.
We define the residual of \feii\ multiplets 37,\,38 (or \feii\,48,\,49) as
the excess of \feii\,37,\,38 (or \feii\,48,\,49) relative to its strength
in the best-fit \feii\, template. A
positive residual means that there is an excess in the \feii\,37,\,38
while a negative value signals an \feii\,48,\,49 excess. 
The \feii\ residual is somewhat
uncertain due to the extinction correction. Since the 
blue light suffers more extinction than the
red light, an over-correction in the
extinction would lead to an over-estimate of the \feii\ 
multiplets 37,\,38 and hence increase the value of 
the \feii\ residual in this multiplets. 
This overall trend  also establishes that
the excess of \feii\ multiplets 48,\,49 relative to
multiplets 37,\,38 cannot be an artifact.
For example, for F00275$-$2859, we have 
adopted E(B$-$V)\,=\,0.41, which results
in a negative \feii\ residual (i.e., there
is an excess in the \feii\ multiplets 48,\,49). If no extinction is
adopted, then the value of the \feii\ residual in
the multiplets 48,\,49 would further increase by 30\%.
We note that while Boroson \& Green (1992) have described
such residuals, they are not as remarkable as in our IR QSO sample.

We list the measurements of significant \feii\ residuals in Table~\ref{tab4}.
We have examined the correlation of the residuals with other parameters.
A correlation was found between the residuals of \feii\ multiplets and
the soft X-ray luminosity. This is interesting because
it ties in with the long-standing puzzle why some NLS1s with
extremely strong \feii\ emission are X-ray luminous, while others
with similar optical spectra
are X-ray weak.
Fig.~\ref{figresiduals} shows the residuals of \feii\ multiplets vs.
the \Lx/\LFIR\ ratio for 14 object with concrete X-ray detections.
We can see from Fig.~\ref{figresiduals} that as the
residual of \feii\ multiplets
increases from minus to positive, the soft X-ray
luminosities of IR QSOs also increase (the S-R correlation coefficient
is 0.44 with the correlation significant of $\calP =0.117$).
The most interesting result from Fig.~\ref{figresiduals} is that objects
with larger \feii\ multiplets 48,\,49 residuals
are all X-ray weak. To check the validity of this result, we measured or estimated the
\feii\ multiplets residual for another
two low-redshift lo-BAL QSOs --- PG~1700+518 and IRAS~14026+4341.
The \feii\ multiplets 48,\,49 residuals are
0.19 and 0.1, respectively. These two objects are shown
in Fig.~\ref{figresiduals} as well
using their upper limits of the soft X-ray luminosity (the soft X-ray
information of PG~1700+518 is from Wang, Brinkmann, \& Bergeronet 1996).
In the same figure, we also indicated the locations of two potential
lo-BAL QSOs, F09427+1929 and IRAS~20036$-$1547, as they have significant
\feii\ multiplets 48,\,49 residuals and extremely weak \OIII. It is
interesting that all four low-redshift lo-BAL QSO and the two 
potential low-BAL IR QSOs are located in the bottom left
of Fig.~\ref{figresiduals}. Furthermore, we also measured the \feii\ multiplets
residual for soft X-ray weak QSOs (Brandt, Laor \& Wills 2000) with available
spectral data. None of them have positive \feii\ multiplets residuals.
It appears that the \feii\ multiplets residual
may be a good criterion to select X-ray weak QSOs or lo-BAL QSOs.

\subsection{Infrared Properties}

The infrared color-color diagram has been used as an important tool
to discriminate starbursts and AGN activities in the nuclear/circumnuclear
regions of galaxies (de Grijp et al. 1985; L\'{\i}pari 1994).
20 IR QSOs in our sample are securely detected in three far-infrared bands
(25, 60 and 100$\mu{\rm m}$).
Fig.~\ref{figcolor} shows the location of these objects
in the infrared color-color diagram,
$\alpha(60, 25)$  versus $\alpha(100, 60)$. Here
$\alpha$($\lambda_1$, $\lambda_2$) =
$-$log(F($\lambda_2$)/F($\lambda_1$))/log($\lambda_2/\lambda_1$),
where the wavelength is in units of $\mu{\rm m}$. In the same diagram,
the power-law and blackbody lines are also indicated.

It is clear from Fig.~\ref{figcolor} that almost all IR QSOs except
F13218+0552 are located between the blackbody and power-law lines.
Moreover, there are a group of IR QSOs clearly located close to the
blackbody line.
This group includes F01572+0009, IR06269$-$0543, F11119+3257, F13218+0552,
F15462$-$0450, F23411+0228 and Mrk~231. We
find that they have either significant \OIII\ blueshifts
(see Table~\ref{tab3}),
or significant \OII\ blueshifts (Mrk~231, see L\'{\i}pari et al. 2002).

The objects close to the power-law line are F18216+6419, F16136+6550,
F12265+0219 (3C~273), F10026+4347, F22454$-$1744 and three additional
objects in
the bottom left of the figure. The first 2 objects are the only classical QSOs
in our sample while  the remaining objects are
all moderate/extremely strong \feii\ emitters with
bright soft X-ray emission (see Table~\ref{tab4}).

\section{Summary and Discussion}

We studied an IR QSO sample
with a total of 25 objects. The sample is compiled
from the QDOT redshift survey,
the 1 Jy ULIRGs survey and from a cross-correlation study of the IRAS Point
Source Catalogue with
the ROSAT All Sky Survey Catalogue. Using the observed optical spectra and
archive data in the infrared and soft X-ray, we
investigated the correlations of the \FeIIHb\ ratio
with the \OIII\ peak, \Hb\ blueshift
and the \Hb\ FWHM. All these parameters are correlated.
We found that soft X-ray weak QSOs, especially lo-BAL QSOs, tend
to  have significant
\feii\ multiplets 48,\,49 (5100--5400~\AA) residuals
(cf. Fig.~\ref{figresiduals}).
The correlation between the \FeIIHb\ residuals and the
\Lx/\LFIR\ ratio, shown in Fig.~\ref{figresiduals}, although somewhat weak,
may be a useful clue for understanding
why some strong/extremely strong \feii\ emitters are X-ray
luminous while others are X-ray quiet.

\subsection{The Outflows}

One of the striking features of the emission lines for IR QSOs is the
blueshift of permitted emission lines or the forbidden \OIII\ line.
As we have removed the \feii\ multiplets carefully surrounding the \Hb\
and \OIIItwo\,, there should not be much \feii\ multiplet contamination to
the \Hb\ and  \OIIItwo\ emission lines.
Moreover, the blueshifts measured from \Hb\ and \Ha\ are similar,
as are the blueshifts measured from the \OIII\ and \OIIIone\ lines
(see Fig. 3). The asymmetries of the \Hb\ and \OIII\ lines cannot be
attributed to contaminations from other emission lines.
Outflows give a plausible explanation for 
the emission line blueshift
(for alternative explanations, see Brandt et al. 2000).
Such emission-line outflows have been found
and discussed extensively
for some NLS1s (Leighly 2000; Christopoulou et al. 1997),
radio galaxies (Tadhunter et al. 2001) and for IR QSOs (L\'{\i}pari 2002).

 From the strong correlation between the \feii\
FWHM and the \Hb\ FWHM,  Boroson \& Green (1992) suggested that
the \feii\ line and permitted emission lines share a common emission region.
If the \Hb\ blueshifts are due to cloud outflows 
in the broad line region, shocks are likely to be produced
in outflows. Such shocks may be responsible for the \feii\ emission, 
as the pure photoionization model fails to explain
the strong \feii\ emission (Collin \& Joly 2000).
The tight correlation between the \Hb\ blueshift and the \FeIIHb\ ratio
can then be understood as they are physically connected through shocks
associated with outflows.
Analogously, the \OIII\ blueshifts could probe the outflows in narrow line emission region.
It is interesting to see that all objects
with large \OIII\ blueshifts are located close to the
blackbody line in the infrared color-color diagram. Hence the locations of
IR QSOs in the infrared color-color diagram
may be related with outflows  in the \OIII\ emission line region.
We caution, however, that there are other explanations
concerning the strong \feii\ emission, especially in NLS1 galaxies (see 
Sulentic, Marziani, \& Dultzin-Hacyan 2000 for a review).

If the emission line blueshifts can indeed be attributed to outflows, then
our sample indicates such outflows are
common for IR QSOs. Our statistics do not however clarify which 
mechanism (central AGN radiative pressure, starbursts
or both) drives these outflows.  High resolution
observations and investigations in the UV, soft X-ray and optical
bands are needed to further explore this unique sample.

\subsection{The Connection of IR QSOs
with Luminous NLS1s and Low-ionization BAL QSOs}

60\% of our IR QSOs satisfy the strict criteria of NLS1s.
For the remaining objects, most are moderate
strong/extremely strong \feii\ emitters. However, 71\% (5/7) of the
extremely strong \feii\
emitters in IR QSO sample fail to meet the strict criterion of NLS1s.
Similarly, a large fraction (44\%) of extremely strong \feii\ emitters in
the sample of V\'{e}ron-Cetty et al. (2001) also have \Hb\ FWHM
larger than 2000~\kms.
As V\'{e}ron-Cetty et al. (2001) pointed out, there should be
a continuous distribution of optical line widths for
QSOs/Seyfert 1s and hence the separation between the broad line
Seyfert 1s and NLS1s at 2000~\kms\ could be arbitrary
(see also Sulentic et al. 2000).
Our results seem to support the above statement and hint that
the strength of optical \feii\ emission line, rather than the \Hb\ FWHM,
may be the most important characteristic of non-classical QSOs.

Much attention has been paid to lo-BAL QSOs recently.
Canalizo \& Stockton (2001b)
studied four lo-BAL QSOs currently known at $z<0.4$ and found
that all four are ULIRGs which
reside in dusty starbursts or post-starbursts. Two of these,
Mrk~231 (i.e., F12540+5708) and IRAS~07598+6508, are in our IR QSO sample.
The main characteristics of lo-BAL QSOs are strong \feii\
emissions, very weak  \OIII\ emissions and low X-ray luminosity.
They are intermediately located between the
power-law and blackbody lines in the infrared color-color diagram
(cf. Fig.~\ref{figcolor}). A recent Chandra survey for BAL QSOs also revealed that
all these four lo-BAL QSOs have high column densities and are
unusually faint in both soft and hard X-ray bands (Gallagher et al. 2002).
These properties are shared by some other
IR QSOs in our sample in addition to Mrk~231 and IRAS~07598+6508.
For example, F00275$-$2859, F09427+1929 and F20036$-$1517
have all the properties of lo-BAL QSOs as described above. We identify them
as potential or more evolved lo-BAL QSOs.
If UV observations establish that they are real lo-BAL QSOs,
the percentage of lo-BAL QSOs in our  sample will be
20\%, similar to the fraction found
by Boroson \& Meyers (1992) for a much smaller IR QSO sample.

Brandt \& Gallagher (2000) argued that the potential physical
connection between luminous
NLS1s and lo-BAL QSOs is their high accretion rate relative to the
Eddington accretion rate. IR QSOs tend to appear in the final merging
phase with their central AGN activity
recently triggered or rejuvenated by the merging activity
(Sanders \& Mirabel 1996; Zheng et al. 1999; Canalizo \& Stockton
 2001a and reference therein). Numerical simulations
show that a large amount of gas flows toward the
center during mergers (e.g., Barnes \& Hernquist 1991), so
it seems plausible that IR QSOs also have high accretion rates.
Such systems may have a greater ability to drive radiative outflows.
The correlations we presented in section 4.2 suggest that the outflow velocity and the physical
condition in the outflow region (inferred from the \Hb\ and
 \OIII\ emission line blueshifts)
influence the \feii\,, infrared and soft X-ray emission
properties. In short, IR QSOs, luminous NLS1s and lo-BAL QSOs
may have some common physical conditions. The difference between them may
be from different viewing
 angles or from different evolution phases.
A careful study of IR QSOs may be important for understanding the
evolution and formation of classical QSOs.

\acknowledgments

We are grateful to Drs. Th. Boller, T.Q. Wang, J.Y. Wei,
L.C. Deng and Mr. Z.H. Shang for helpful discussions,
and to the BATC members for advice on data deduction.
Particular thanks are due to Drs T. Boroson \& R. Green for kindly providing their \feii\ template
and their dataset of QSOs in PG92. We also thank the anonymous referee 
and Dr. Richard James for
constructive comments that improved the paper.
This project was supported by NSF of China and NKBRSF G19990754.

\clearpage

\begin{figure}
\epsscale{0.9}
\plotone{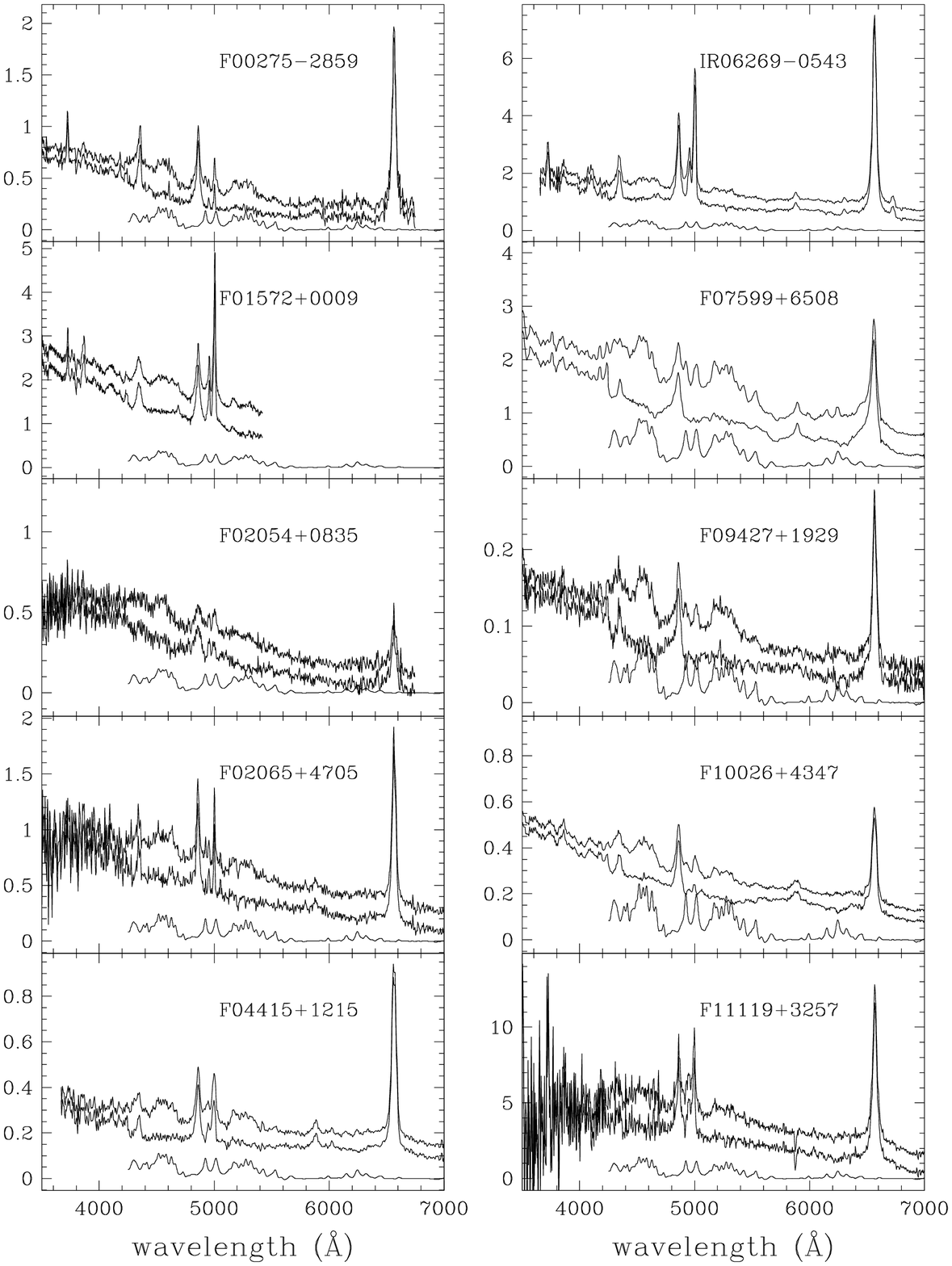}
\caption{Dereddened spectra (top curves), \feii-subtracted spectra
(middle curves) and \feii\ spectra (bottom curves) of 25 IR QSOs,
shown in increasing order of right ascension. The
\feii-subtracted spectra have been shifted downwards for clarity.
The vertical axis shows the observed flux in units of
10$^{-15}$~erg\,cm$^{-2}$\,s$^{-1}$\,\AA$^{-1}$.  The horizontal axis
is wavelength in \AA\ in the rest frame.
\label{figspectra}}
\end{figure}

\setcounter{figure}{0}
\begin{figure}
\epsscale{0.9}
\plotone{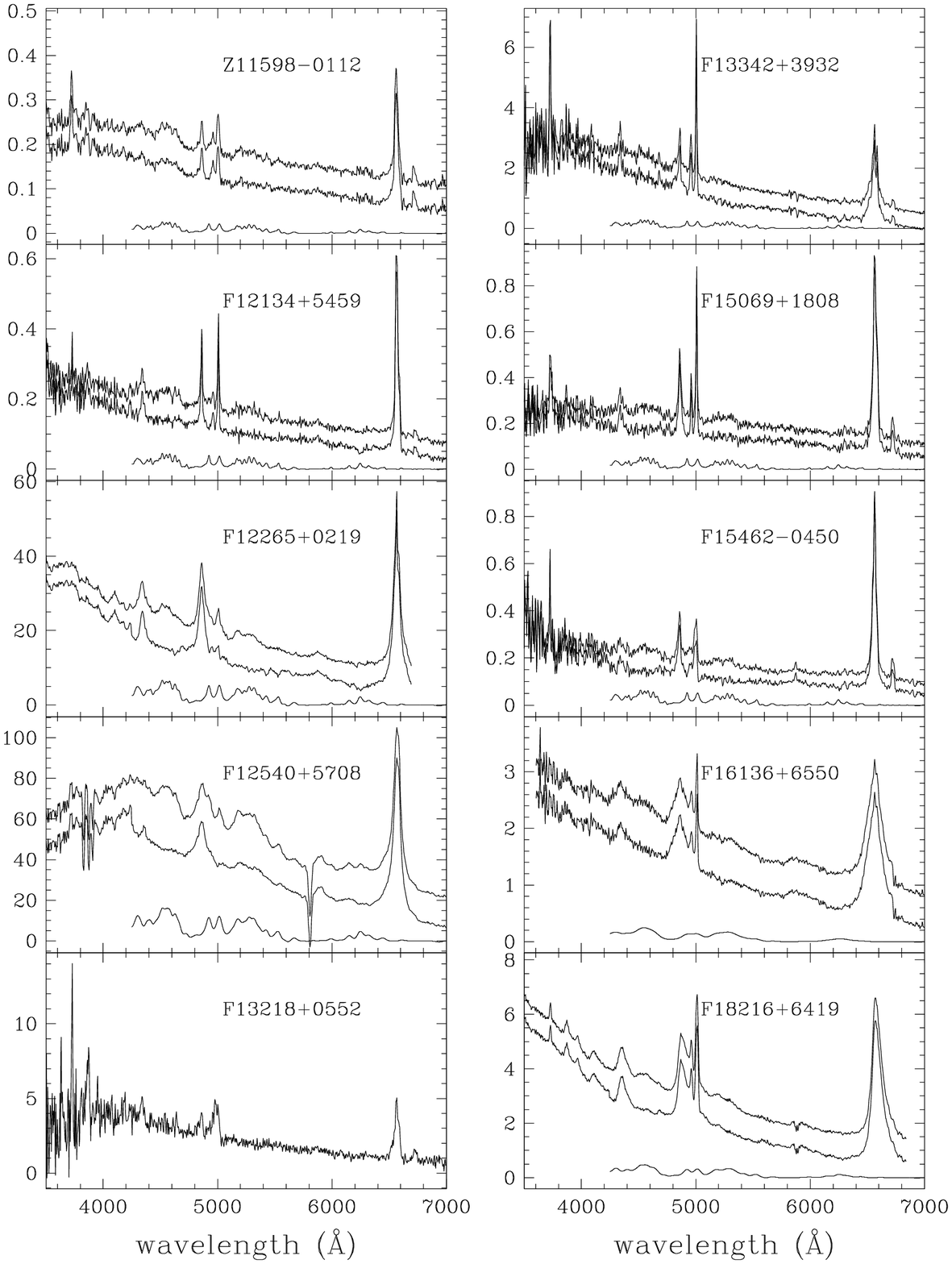}
\caption{Continued.}
\end{figure}

\setcounter{figure}{0}
\begin{figure}
\epsscale{0.9}
\plotone{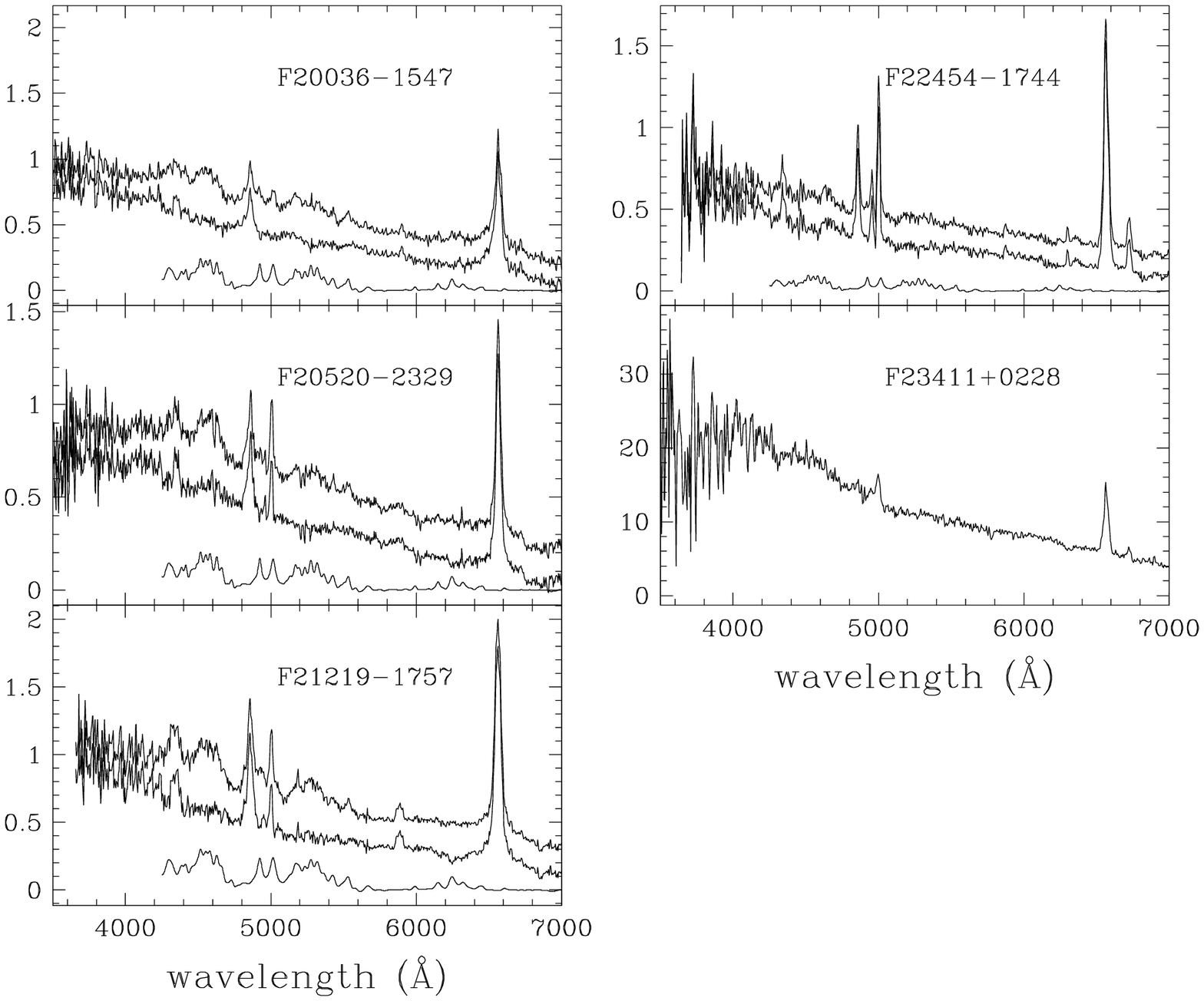}
\caption{Continued.}
\end{figure}

\setcounter{figure}{1}
\begin{figure}
\plotone{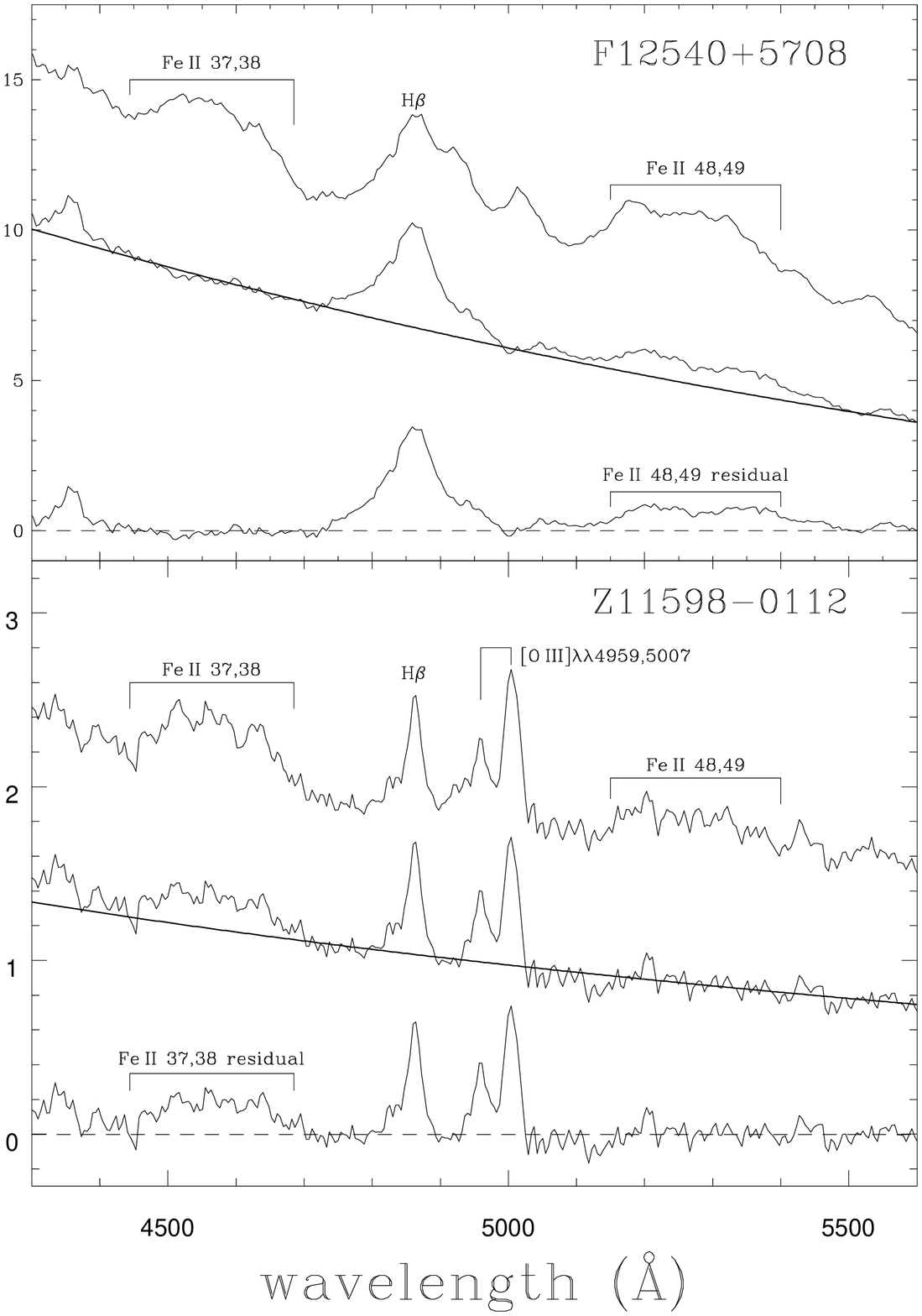}
\caption{
Illustration of significant \feii\ multiplets residuals. Top panel:
F12540+5708 shows the \feii\ multiplets 48,\,49 residual compared with the
\feii\ Boroson \& Green (1992) template. Bottom panel: Z11598$-$0112
shows \feii\ multiplets 37,\,38 residual. In each panel, the top curve
is the dereddened observed spectrum between 4300-5100~\AA. The middle
one show the \feii-subtracted spectrum (with an arbitrary offset)
with the fitted low-order polynomial continuum shown as a thick solid
line.  The excess of \feii\ emission lines is illustrated in the
continuum-subtracted bottom curve.  \label{FeIIresiduals}
}
\end{figure}

\begin{figure}
\plotone{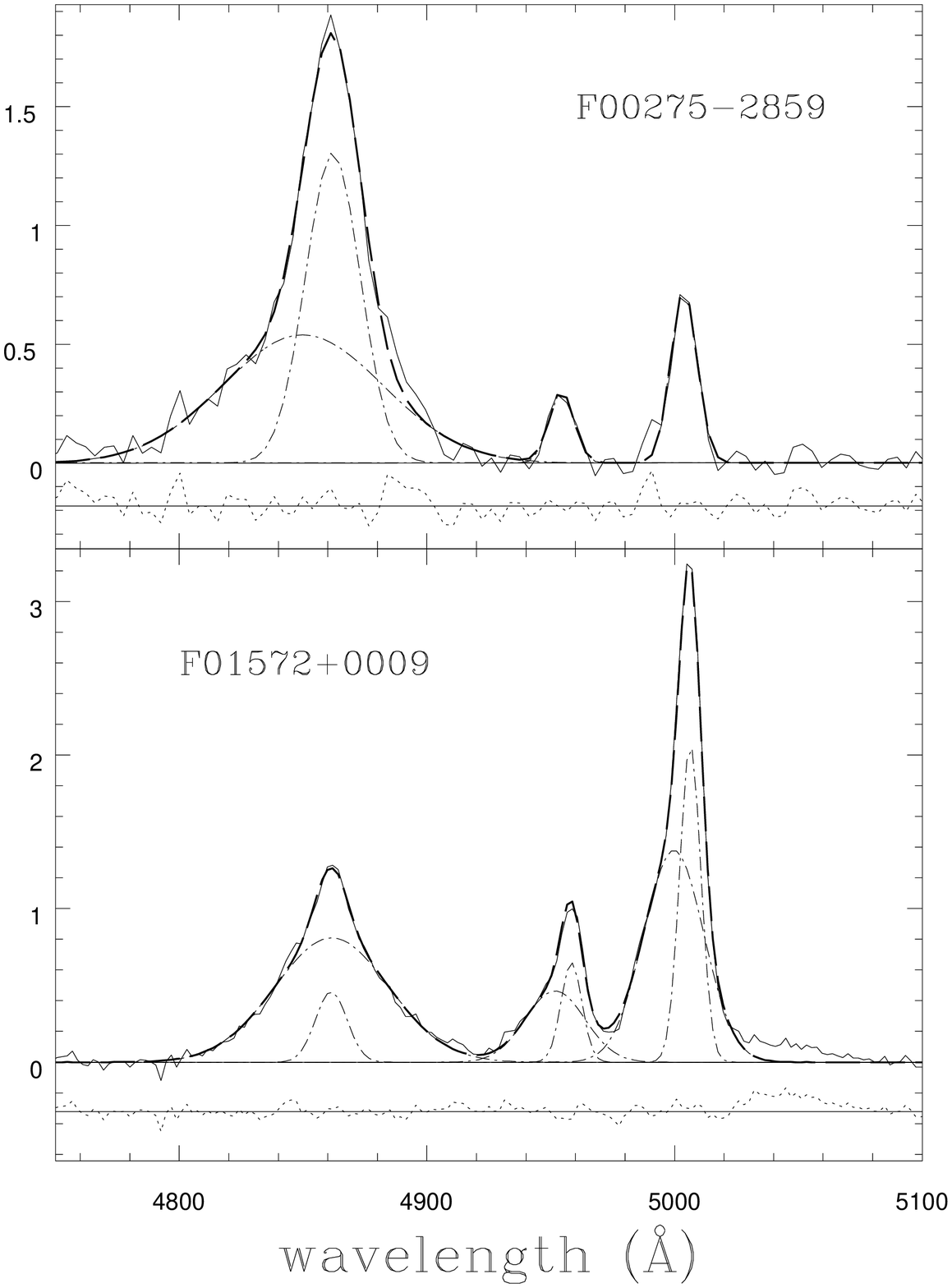}
\caption{Illustration of the blueshifted \Hb\ and \OIIItwo\ lines.
Top panel: For F00275$-$2859, \Hb\ can be well fitted by a
sum of two Gaussian components with the broad component blueshifted
by 750~\kms\ relative to the narrow one. Bottom panel: For
F01572+0009, \OIIItwo\ shows a narrow component in addition to a
blueshifted broad component. The solid line is the observed profile.
The dashed line shows the fitted profile. Each fitted component is
shown by a dot-dashed line and the dotted line illustrates the
residual of fitting.
\label{figtemplate}}
\end{figure}

\begin{figure}
\plotone{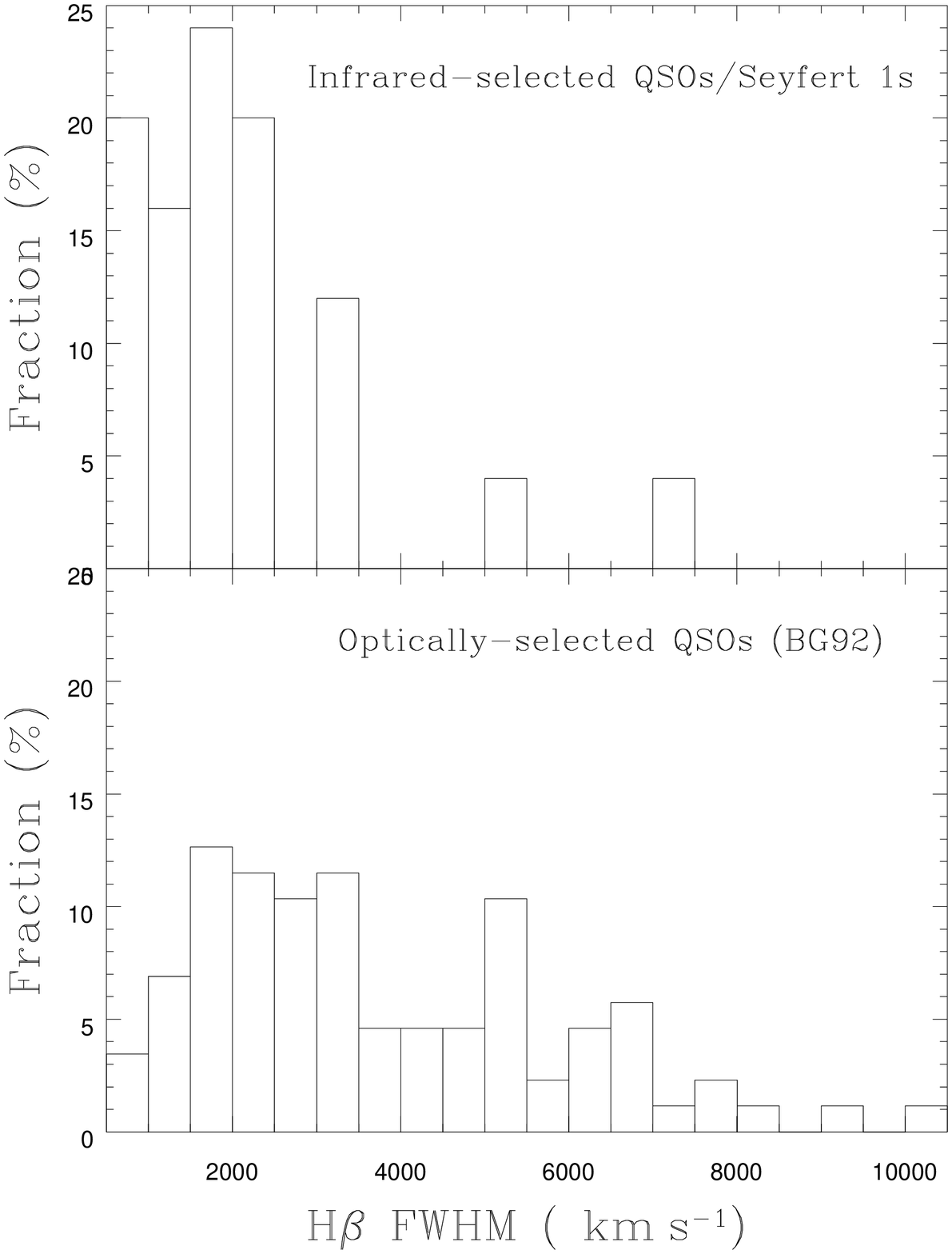}
\caption{Histograms of the \Hb\ FWHM for IR QSOs (top panel) and 
for the Boroson \& Green sample (bottom panel),
which includes 87 optically selected bright QSOs.  \label{fighistogram1}}
\end{figure}

\begin{figure}
\plotone{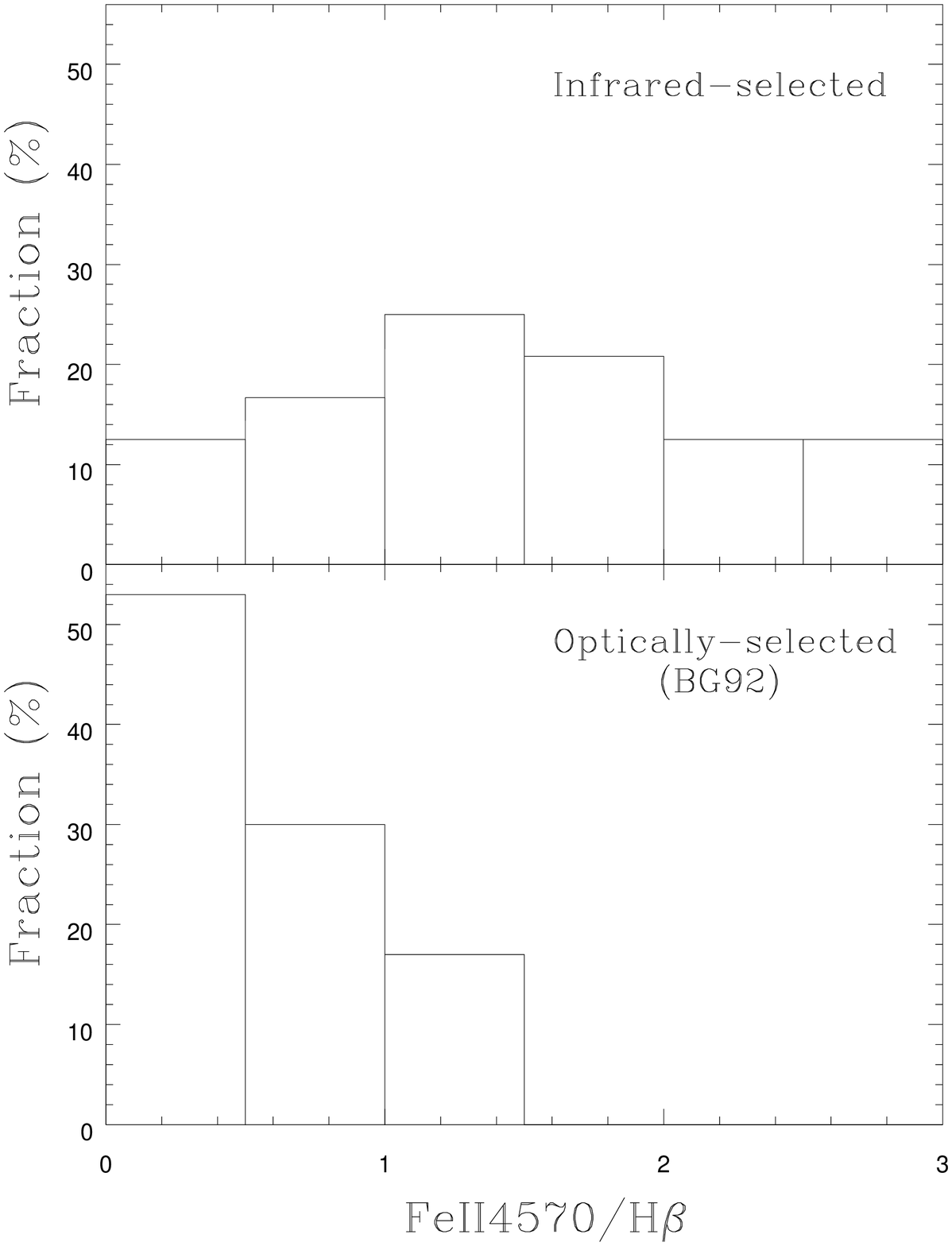}
\caption{Histograms of the \FeIIHb\ ratio for IR QSOs (top panel) and
the BG92 sample (bottom panel).
\label{fighistogram2}}
\end{figure}

\begin{figure}
\plotone{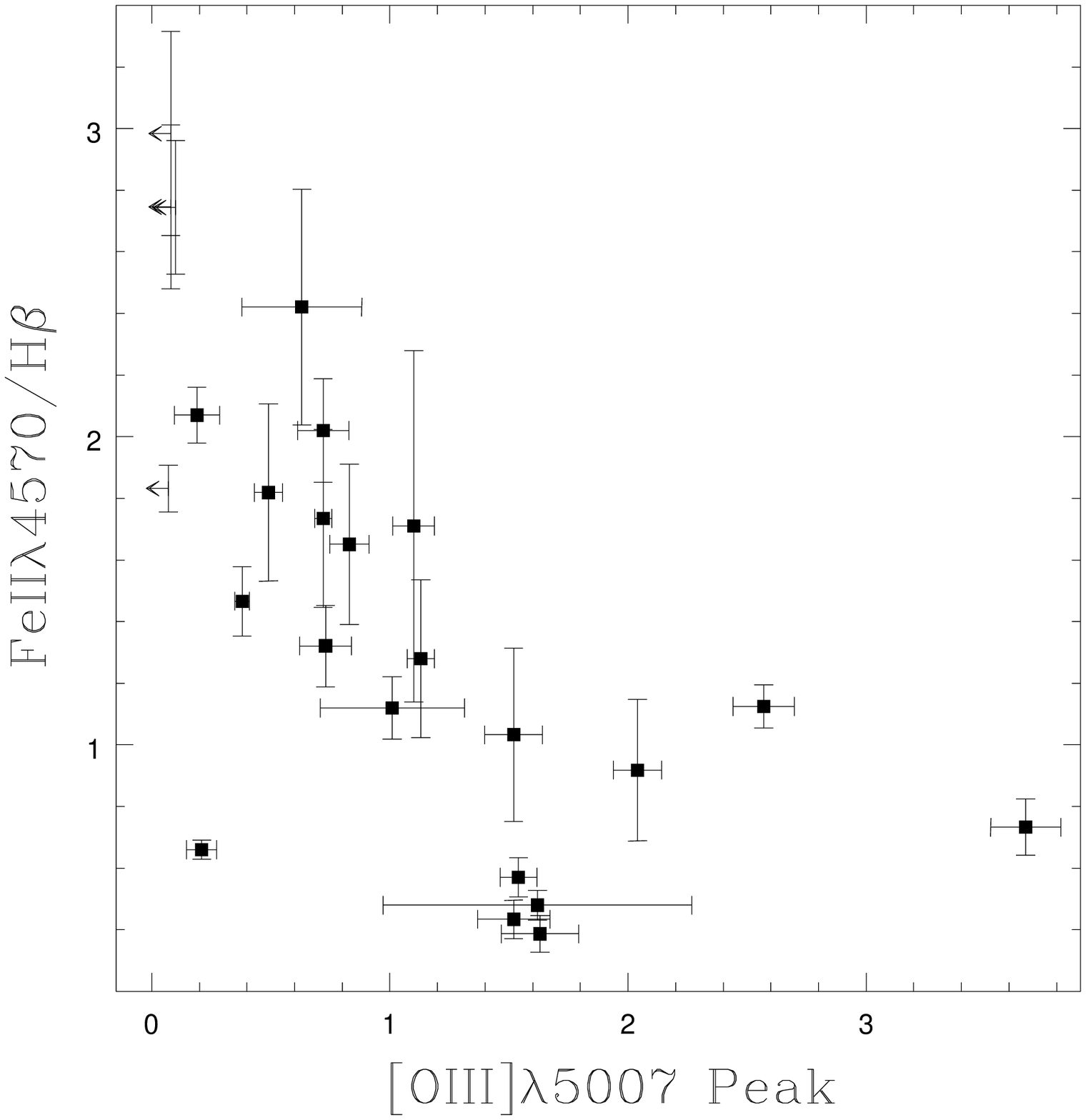}
\caption{\FeIIHb\ vs. the ratio of the peak height
of \OIII\ to that of \Hb. Upper limits for the non-detections
are indicated by arrows.
For the detected objects, the Spearman Rank-order correlation
coefficient is $-$0.78 with $\calP =6.1 \times 10^{-6}$, i.e.,
the probability for the null hypothesis of no correlation
between these two parameters being true is $6.1 \times 10^{-6}$.
\label{figewoiii_feiihb}}
\end{figure}

\begin{figure}
\plotone{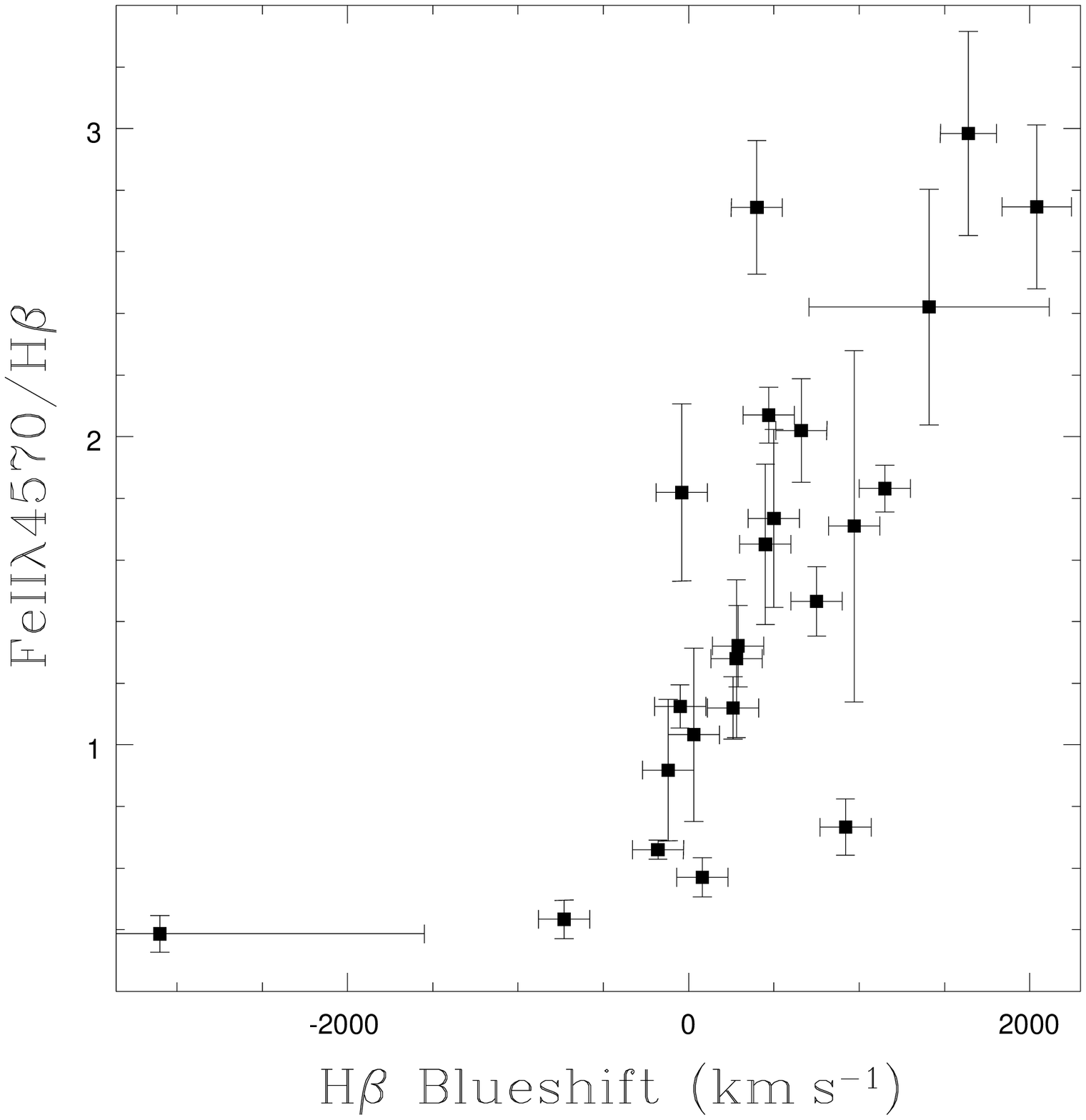}
\caption{\FeIIHb\ vs. the \Hb\ blueshift, which is defined as the blueshift
of the \Hb\ broad component relative to the \Hb\ narrow component in unit
of \kms. A negative value indicates a redshift. For 23 objects, the
correlation coefficient is 0.76 with $\calP =2.7 \times 10^{-5}$.
\label{figbs}}
\end{figure}

\begin{figure}
\plotone{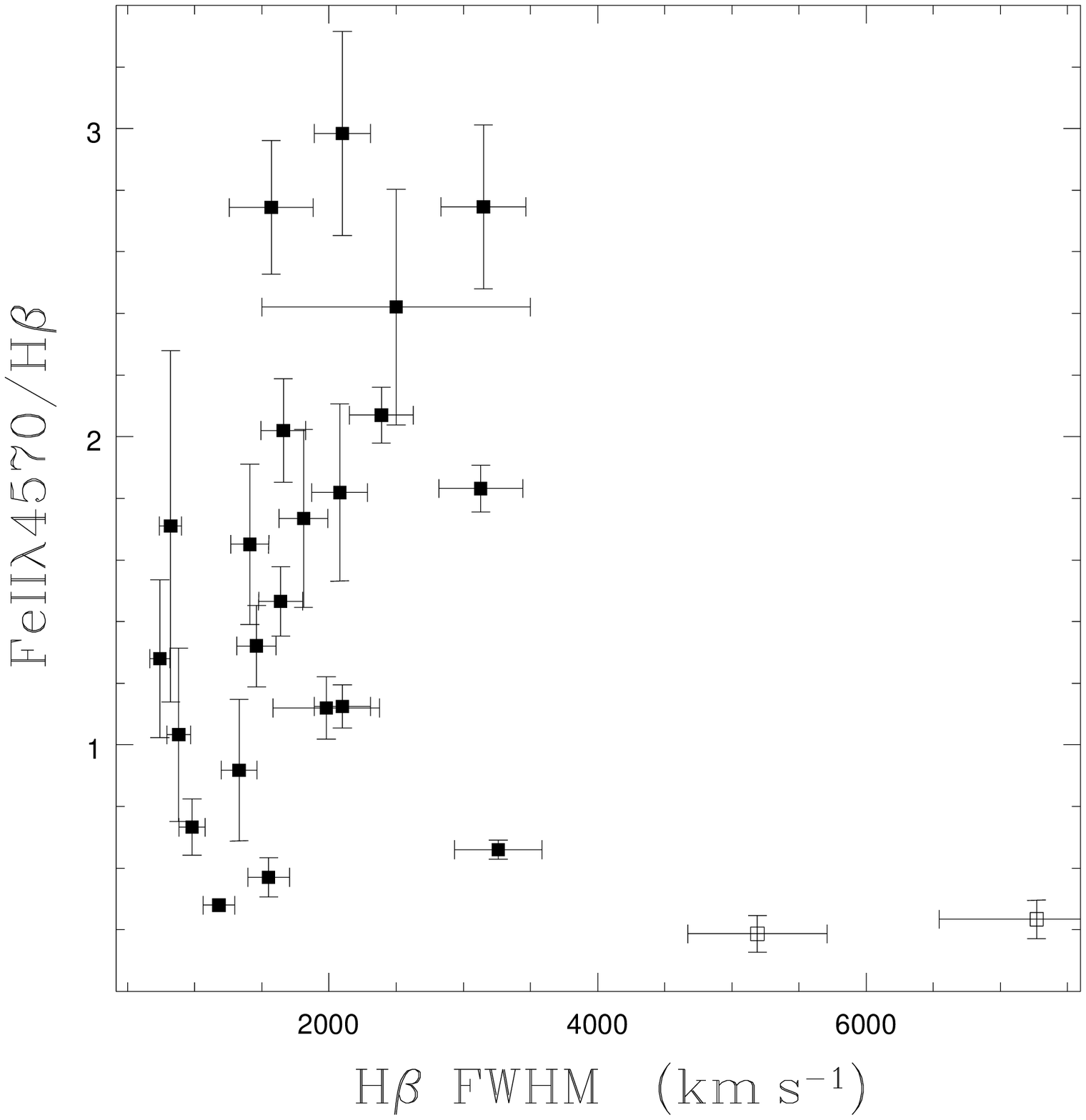}
\caption{
\FeIIHb\ vs. \Hb\ FWHM. The 22 Objects with 
\Hb\ FWHM less than 4000~\kms are shown as filled
squares while the two objects with
\Hb\ FWHM greater than 4000~\kms are shown with open squares.
The Spearman Rank-order correlation coefficient for the objects 
with FWHM smaller than 3000~\kms (4000~\kms) 
is 0.47 (0.58) corresponding to $\calP = 2.8 \times 10^{-2}$
($\calP = 9.9 \times 10^{-3}$).
No correlation is apparent if we include all objects.  
Note that F23411+0228 is excluded due to its low S/N  
spectrum.  
\label{figfwhm_feiihb}}
\end{figure}

\begin{figure}
\plotone{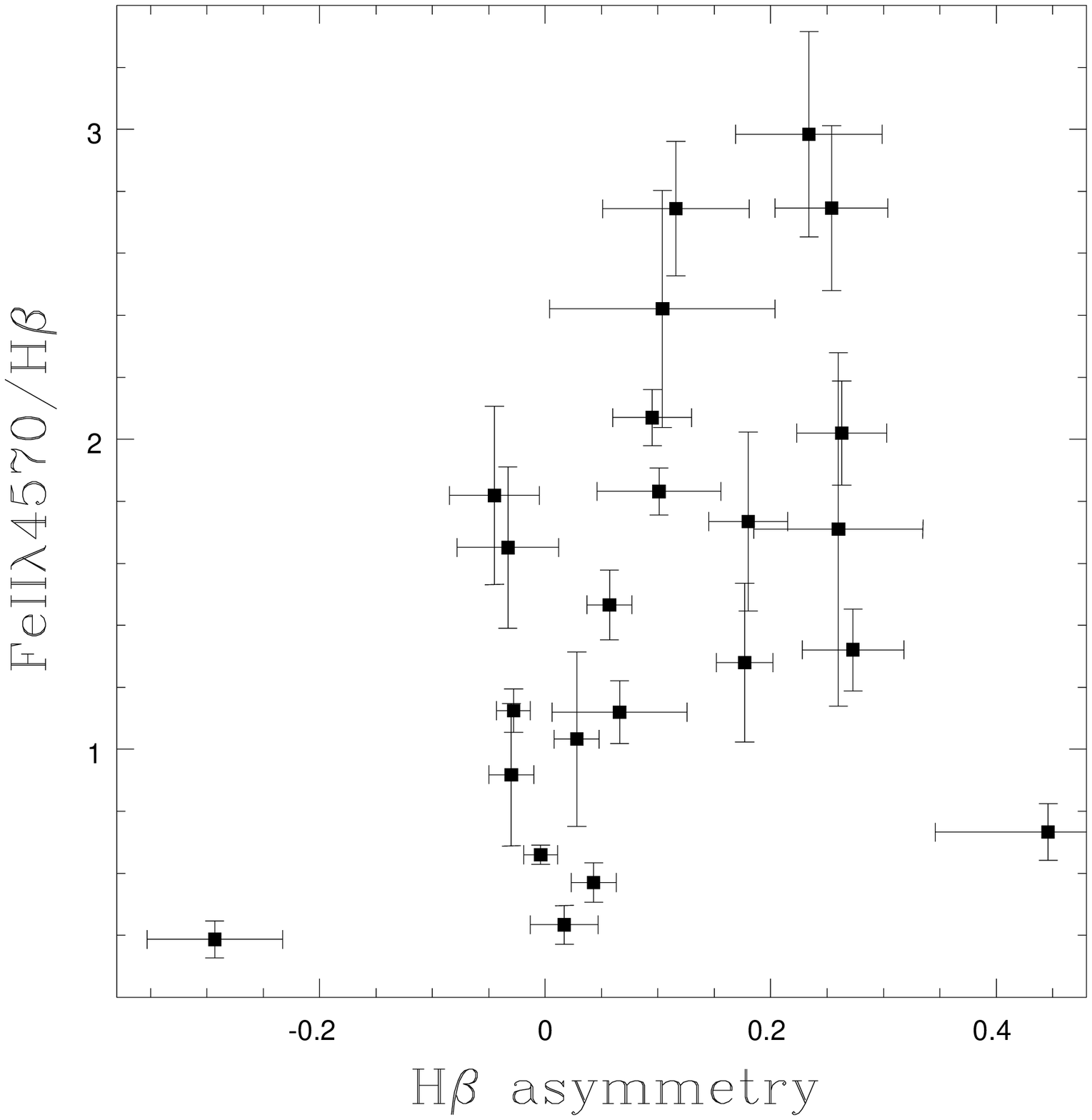}
\caption{\FeIIHb\ vs. the \Hb\ asymmetry index defined by de Robertis
(1985, cf. eq. 1). The correlation coefficient is 0.46
corresponding to  $\calP =2.8 \times 10^{-2}$.
\label{figasyfeii}}
\end{figure}

\begin{figure}
\plotone{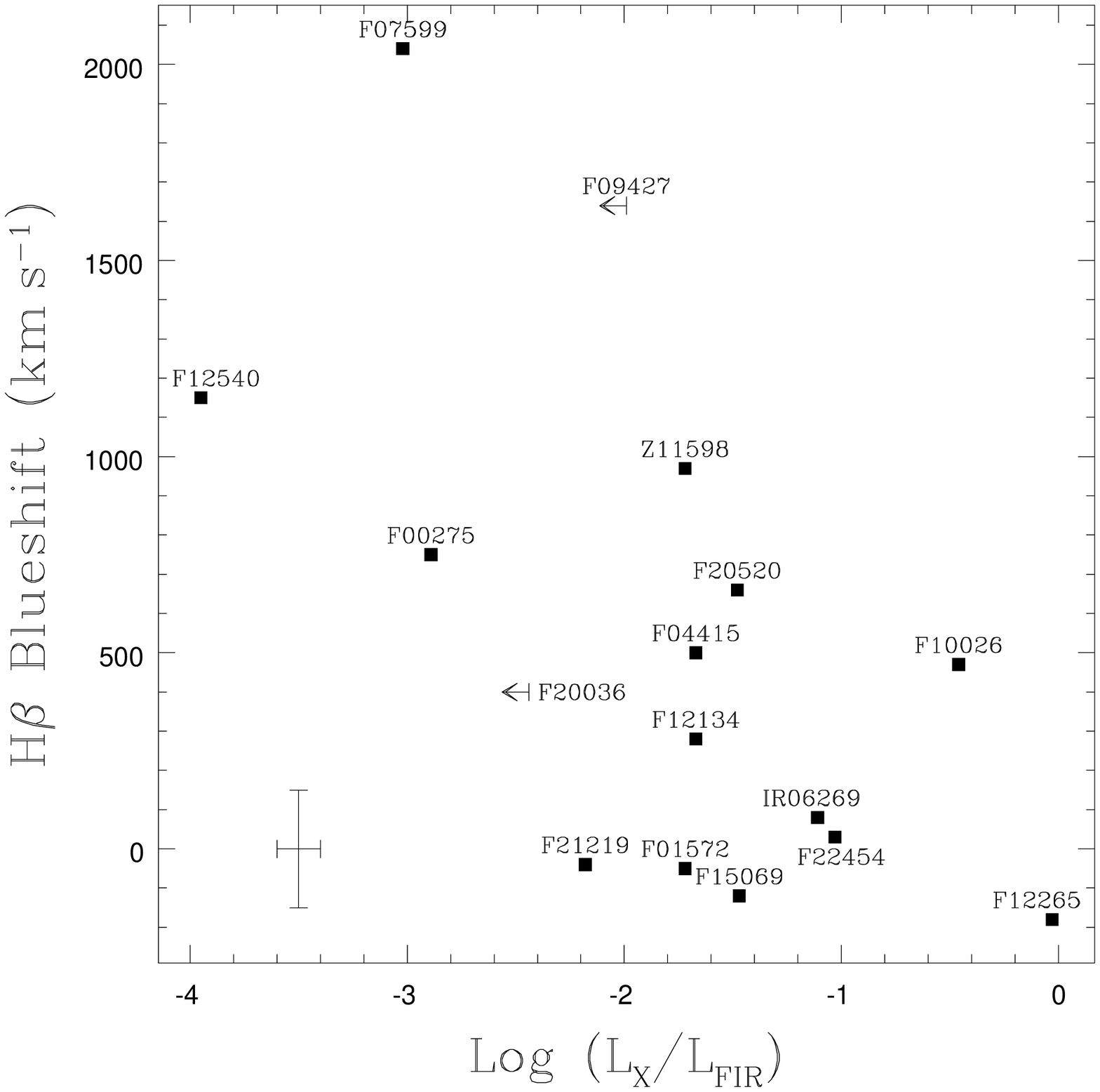}
\caption{\Hb\ blueshift vs. soft X-ray luminosity for 14 objects (filled
squares) detected by the ROSAT All Sky Survey or Pointings.
The correlation coefficient for the detected objects is $-$0.61 
with $\calP =2.0 \times 10^{-2}$.
Additionally, two potential lo-BAL QSOs are marked by arrows with
their upper limits of X-ray luminosity (see text). Typical error
bars are indicated in the lower-left.
\label{figxraybs}}
\end{figure}

\begin{figure}
\plotone{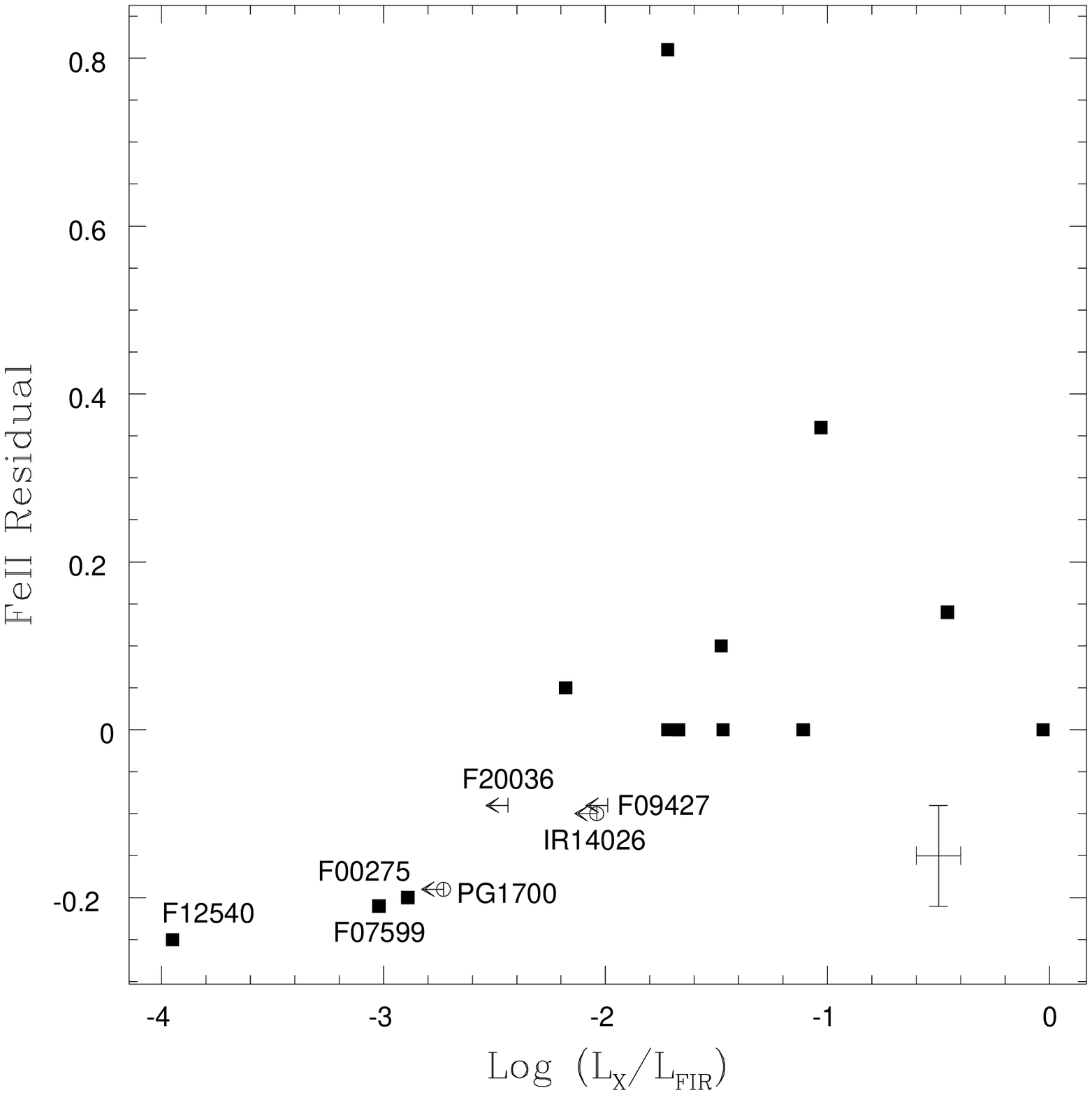}
\caption{\feii\ residual vs. \Lx/$L_{\rm FIR}$.
The \feii\ residual is defined in the text (see section 4.3).
The arrows with a circle indicate upper limits for additional lo-BAL QSOs while
those without a circle are for potential lo-BAL QSOs. For the 14 ROSAT detected
objects (filled squares), the correlation coefficient is 0.44 
with $\calP =0.117$. Typical error bars are indicated in the lower-right.
\label{figresiduals}}
\end{figure}

\begin{figure}
\plotone{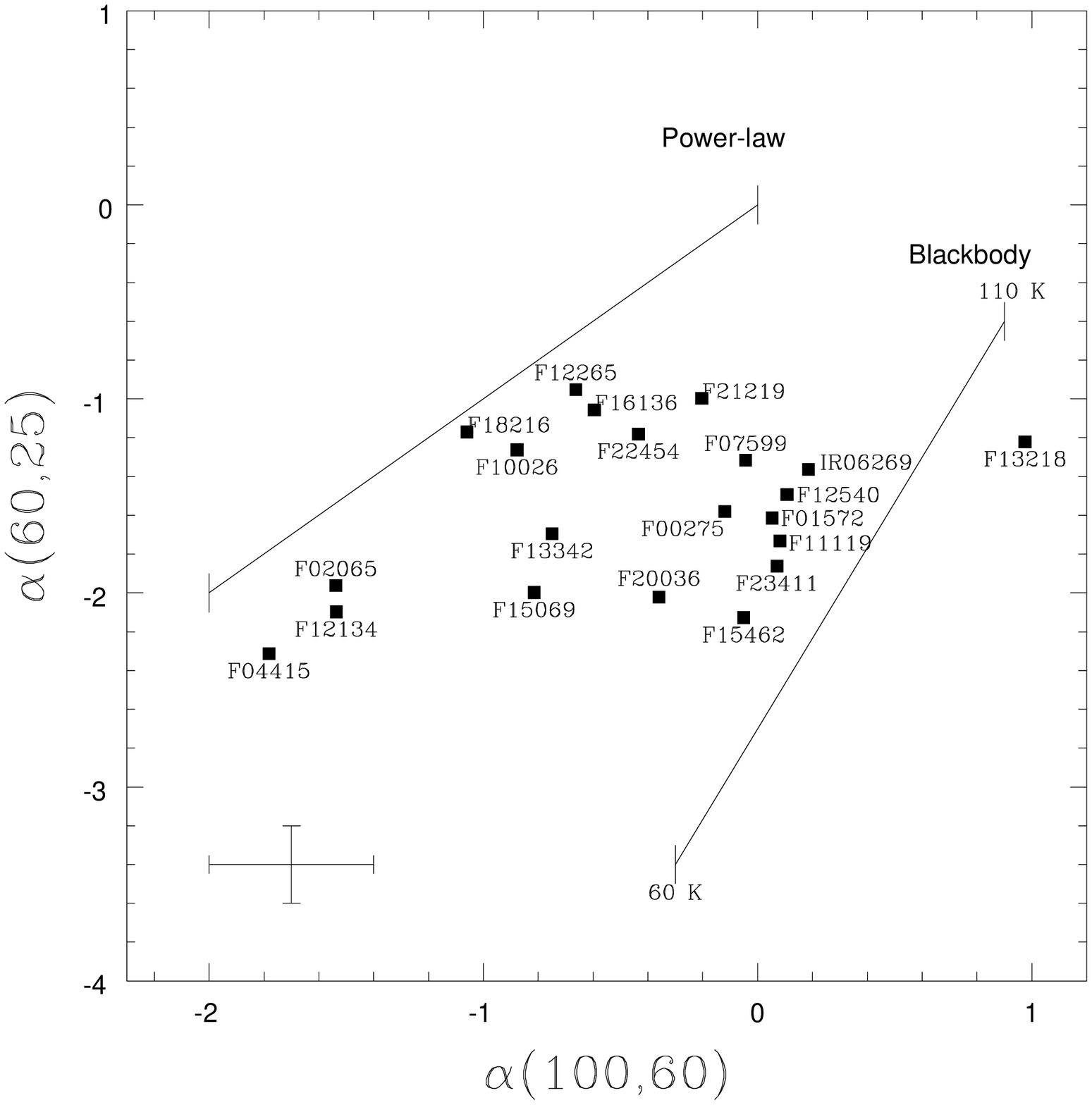}
\caption{IRAS color-color diagram. Our sample galaxies fall in the
region between the
power-law and blackbody lines. The objects close to the
blackbody line all have significant \OIII\ blueshifts.
\label{figcolor}}
\end{figure}

\clearpage

\begin{table}
\begin{center}
\caption{Sample of IR QSOs Selected from ULIRGs. \label{tab1}}
\begin{tabular}{cccccrcl}
\noalign{\smallskip}
\tableline
\tableline
\noalign{\smallskip}
IRAS Name    & R.A.  &  Decl. &
Log$\left (L_{\rm FIR} \over L_\odot \right)$ &
Log$\left (L_{\rm IR} \over L_\odot \right)$ &
Log$\left (L_{\rm X} \over L_{\rm IR} \right)$ & Redshift &  Ref. \\
(1) & (2) & (3) & (4) & (5) & {\parbox[t]{15mm}{\centering (6)}} & (7) & (8) \\
\noalign{\smallskip}
\tableline
\noalign{\smallskip}

F00275$-$2859 &  00 30 04.2 &$-$28 42 24.6 & 12.64 & 12.90 &   $-2.89$ & 0.279  &  (1) \\
F01572$+$0009 &  01 59 50.2 &  +00 23 42.2 & 12.65 & 12.85 &   $-1.72$ & 0.163  &  (2) \\
F02054$+$0835 &  02 08 06.8 &  +08 50 05.2 & 12.97 & 13.29 &  $<-2.17$ & 0.345  &  (1) \\
F02065$+$4705 &  02 09 45.8 &  +47 19 43.2 & 12.27 & 12.45 &  $<-2.47$ & 0.132  &  (1) \\
F04415$+$1215 &  04 44 28.8 &  +12 21 13.1 & 12.41 & 12.53 &   $-1.67$ & 0.089  &  (3) \\
IR06269$-$0543 &  06 29 24.7 &$-$05 45 26.0 & 12.49 & 12.74 &   $-1.11$ & 0.117  &  (3) \\
F07599$+$6508 &  08 04 30.4 &  +64 59 53.3 & 12.45 & 12.77 &   $-3.02$ & 0.148  &  (2) \\
F09427$+$1929 &  09 45 27.6 &  +19 15 42.1 & 12.61 & 12.90 &  $<-1.99$ & 0.284  &  (4) \\
F10026$+$4347 &  10 05 41.8 &  +43 32 41.6 & 12.20 & 12.54 &   $-0.46$ & 0.178  &  (1) \\
F11119$+$3257 &  11 14 38.8 &  +32 41 34.7 & 12.64 & 12.88 &  $<-2.40$ & 0.189  &  (2) \\
Z11598$-$0112 &  12 02 26.6 &$-$01 29 15.3 & 11.91 & 12.43 &   $-1.72$ & 0.151  &  (2,3)\\
F12134$+$5459 &  12 15 49.3 &  +54 42 24.6 & 12.17 & 12.36 &   $-1.67$ & 0.150  &  (3) \\
F12265$+$0219 &  12 29 06.6 &  +02 03 09.0 & 12.65 & 13.04 &   $-0.03$ & 0.158  &  (2,3)\\
F12540$+$5708 &  12 56 13.9 &  +56 52 24.6 & 12.60 & 12.82 &   $-3.95$ & 0.042  &  (1,2)\\
F13218$+$0552 &  13 24 19.9 &  +05 37 04.6 & 12.53 & 12.94 &  $<-2.23$ & 0.205  &  (2) \\
F13342$+$3932 &  13 36 24.0 &  +39 17 32.2 & 12.49 & 12.72 &  $<-2.30$ & 0.179  &  (2) \\
F15069$+$1808 &  15 09 13.7 &  +17 57 11.0 & 12.24 & 12.47 &   $-1.47$ & 0.171  &  (3) \\
F15462$-$0450 &  15 48 56.8 &$-$04 59 33.5 & 12.35 & 12.50 &  $<-2.68$ & 0.101  &  (2) \\
F16136$+$6550 &  16 13 57.1 &  +65 43 11.0 & 11.92 & 12.24 &   $-0.21$ & 0.129  &  (3) \\
F18216$+$6419 &  18 21 57.1 &  +64 20 37.4 & 13.02 & 13.34 &   $-0.73$ & 0.297  &  (3) \\
F20036$-$1547 &  20 06 31.9 &$-$15 39 05.8 & 12.70 & 12.89 &  $<-2.44$ & 0.193  &  (1) \\
F20520$-$2329 &  20 54 57.3 &$-$23 18 24.8 & 12.52 & 12.77 &   $-1.48$ & 0.206  &  (3) \\
F21219$-$1757 &  21 24 41.6 &$-$17 44 45.3 & 12.02 & 12.39 &   $-2.18$ & 0.113  &  (2) \\
F22454$-$1744 &  22 48 04.1 &$-$17 28 28.5 & 11.94 & 12.37 &   $-1.03$ & 0.117  &  (3) \\
F23411$+$0228 &  23 43 39.7 &  +02 45 05.7 & 12.14 & 12.34 &   $-1.98$ & 0.091  &  (3) \\
\noalign{\smallskip}
\tableline
\noalign{\smallskip}
\end{tabular}
\tablecomments{
 The Prefix of the object name indicates the origin of IRAS fluxes. `F' refers
 to the IRAS Faint Source Catalogue and `Z' means the Faint Source
 Reject File (see Moshir et al. 1992). For IR06269$-$0543, the IRAS
 fluxes come from the IRAS Point Source Catalogue.  Units of right
 ascension are hours, minutes, and seconds, and units of declination are
 degrees, arcminutes, and arcseconds (J2000.0).
 Col.(4) \& (5): far-infrared \& infrared luminosity, calculated following Sanders \& Mirabel (1996); Col.(6): Soft X-ray luminosity (0.2--2.4 keV) normalized to far-infrared luminosity; Col.(7): redshift, taken from the references.}
\tablerefs{(1) Lawrence et al. 1999; (2) Kim \& Sanders 1998; (3) Moran et al. 1996; (4) Zheng et al. 1999.}
\end{center}
\end{table}

\clearpage

\begin{table}
\begin{center}
\caption{Journal of Observations.\label{tab2}}
\begin{tabular}{lcccc}
\noalign{\smallskip}
\tableline
\tableline
\noalign{\smallskip}
          &        & Exp.Time &  Slit   &  Seeing \\
IRAS Name &   Date & (second) &(arcsec) &(arcsec) \\
\noalign{\smallskip}
\tableline
\noalign{\smallskip}
F00275$-$2859  & 1999 Nov 07  & 3600  &  3.0  &  3.0 \\
F01572$+$0009  & 1996 Nov 17  & 3600  &  3.0  &  3.0 \\
F02054$+$0835  & 1998 Oct 22  & 2700  &  3.0  &  2.1 \\
F02065$+$4705  & 1999 Feb 20  & 1800  &  2.2  &  1.5 \\
F04415$+$1215  & 1998 Oct 18  & 3600  &  3.0  &  3.5 \\
IR06269$-$0543  & 1998 Oct 23  & 1800  &  3.0  &  2.1 \\
F07599$+$6508  & 1997 Mar 16  & 3600  &  3.0  &  1.5 \\
F09427$+$1929  & 1998 Dec 20  & 3600  &  2.5  &  1.5 \\
F10026$+$4347  & 1999 Feb 22  & 3600  &  2.2  &  1.5 \\
F11119$+$3257  & 1999 Feb 21  & 2400  &  2.2  &  1.5 \\
Z11598$-$0112  & 1997 Mar 12  & 5400  &  3.0  &  3.0 \\
F12134$+$5459  & 1999 Feb 22  & 2700  &  2.2  &  1.5 \\
F12265$+$0219  & 1995 Mar 13  & 600   &  3.0  &  3.0 \\
F12540$+$5708  & 1997 Mar 16  & 1200  &  3.0  &  1.5 \\
F13218$+$0552  & 1999 Feb 21  & 3600  &  2.2  &  2.0 \\
F13342$+$3932  & 1999 Feb 22  & 1500  &  2.2  &  1.5 \\
F15069$+$1808  & 1999 Feb 22  & 1200  &  2.2  &  1.5 \\
F15462$-$0450  & 1997 Apr 11  & 3600  &  3.0  &  2.0 \\
F16136$+$6550  & 1998 Oct 20  & 1800  &  3.0  &  2.1 \\
F18216$+$6419  & 1998 Oct 17  & 2700  &  2.5  &  3.5 \\
F20036$-$1547  & 1997 Oct 03  & 2400  &  2.5  &  3.5 \\
F20520$-$2329  & 1998 Oct 22  & 2700  &  3.0  &  2.1 \\
F21219$-$1757  & 1998 Oct 20  & 2700  &  3.0  &  2.1 \\
F22454$-$1744  & 1998 Oct 20  & 2700  &  3.0  &  2.1 \\
F23411$+$0228  & 1997 Oct 04  & 4800  &  2.5  &  3.5 \\
\noalign{\smallskip}
\tableline
\noalign{\smallskip}
\end{tabular}
\end{center}
\end{table}

\clearpage

\begin{table}
\begin{center}
\caption{Profile properties of \Hb\ and \OIII. \label{tab3}}
\begin{tabular}{cccrccc}
\noalign{\smallskip}
\tableline
\tableline
\noalign{\smallskip}
  & \multicolumn{3}{c}{H$\beta$} & & \multicolumn{2}{c}{[\ion{O}{3}]$\lambda$5007} \\
\cline{2-4} \cline{6-7} \\
         & FWHM &  Blueshift & asy & & FWHM & Blueshift\tablenotemark{a} \\
IRAS Name & (\kms) &  (\kms)      &     & & (\kms) & (\kms) \\
(1) & {\parbox[t]{15mm}{\centering (2)}} & (3) & (4) & & {\parbox[t]{15mm}{\centering (5)}} & (6) \\
\noalign{\smallskip}
\tableline
\noalign{\smallskip}
       F00275$-$2859 & 1640 &    750 &   0.057 &&  710 &  \\
         F01572+0009 & 2100 &  $-$50 &$-$0.028 &&  680 &510 \\
         F02054+0835 & 2500;&  1410; &   0.104 && 1300; &  \\
         F02065+4705 & 1410 &   450: &$-$0.033 &&  500 & \\
         F04415+1215 & 1810 &  510   &   0.180 && 1240 & \\
      IR06269$-$0543 & 1550 &   80   &   0.043 && 1210 &550  \\
         F07599+6508 & 3150 & 2030   &   0.254 &&  ... & \\
         F09427+1929 & 2100 & 1640   &   0.234 &&  ... & \\
         F10026+4347 & 2390 &  470   &   0.095 &&  810 & \\
         F11119+3257 & 1980:&  260:  &   0.066 && 1480: & 950;  \\
       Z11598$-$0112 &  820 &  970   &   0.260 &&  550 & \\
         F12134+5459 &  740 &  280   &   0.177 &&  780 &  \\
F12265+0219\tablenotemark{b}&3260&$-$180: &$-$0.004 && 1330:&1020 \\
         F12540+5708 & 3130 & 1150   &   0.101 &&  ... & \\
         F13218+0552 &1180\tablenotemark{c}&  ... & ...     &&  1530; &500; \\
         F13342+3932 &  980 &  920   &   0.446 &&  540  & \\
         F15069+1808 & 1330 & $-$120 &$-$0.030 &&  630  & \\
       F15462$-$0450 & 1460 &   290: &   0.273 && 1720:&1110: \\
F16136+6550\tablenotemark{b}& 7270 & $-$730: &   0.017 &&  920 & 780 \\
F18216+6419\tablenotemark{b} & 5190  & $-$3100; &$-$0.293 && 1360 & \\
       F20036$-$1547 & 1570: & 400:  &   0.116 && ...  & \\
       F20520$-$2329 & 1660 &  660   &   0.263 &&  300 &560: \\
       F21219$-$1757 & 2080 & $-$40  &$-$0.045 && 1360 & 460:\\
       F22454$-$1744 &  880 &   30   &   0.028 &&  790 & \\
   F23411+0228 &  970\tablenotemark{c} &     ... &    ...  && 1390; & 500; \\
\noalign{\smallskip}
\tableline
\noalign{\smallskip}
\end{tabular}
\tablenotetext{a}{For sources that have an obvious asymmetric \OIIItwo\ profile.}
\tablenotetext{b}{Cases with  broad \Hb\ line; they could be fitted
satisfactorily with three Gaussian components.}
\tablenotetext{c}{FWHM of \Ha.}
\tablecomments{
The uncertainty on the FWHM is typically of order 10\%
($\sim$\,200~\,\kms). Colons (:) indicate values with a relative
uncertainty of 20\%. Semicolons (;) indicate values with a relative
uncertainty 30\%--40\%.  The typical uncertainty of the blueshift is
$\la 150$~\,\kms. Colons (:) indicate values with a relative
uncertainty of 30\%. Semicolons (;) indicate values with a relative
uncertainty of larger than 50\%.}
\end{center}
\end{table}

\clearpage

\begin{table}
\scriptsize
\begin{center}
\caption{Emission Lines Properties.\label{tab4}}
\begin{tabular}{ccccrrrcccccr}
\noalign{\smallskip}
\tableline
\tableline
\noalign{\smallskip}

 & & & {\rm Fe\,{II}} & $\lambda$5007 & & & & \Ha\ & \Hb\ &{\rm Fe\,{II}}  & {\rm [O\,{III}]}& {\rm [O\,{II}]} \\
IRAS Name & E(B$-$V) & ${\rm Fe\,{II} \over H\beta}$ & residual & Peak &${\rm [O\,{III}]\over H\beta}$\, &${\rm [O\,{II}] \over H\beta}$\, &${\rm [N\,{II}] \over H\alpha}$ & EW & EW & EW & EW & EW \\
          &          & &          &               & & & & (\AA)&(\AA) &(\AA) &(\AA) &(\AA) \\
(1) & (2) & (3) &(4) & (5) & (6) & (7) & (8) & (9) & (10) & (11) & (12) & (13) \\
\noalign{\smallskip}
\tableline
\noalign{\smallskip}

F00275$-$2859 & 0.41 & 1.47 & $-$0.20 &   0.38& 0.14 & 0.23 & $<$0.01 & 453.0 & 75.6 & 110.8 & 10.7 & 7.5          \\
  F01572+0009 & 0.00 & 1.13 & 0.0 &   2.57& 1.05 & 0.11 & $<$0.01 &       & 43.5 &  49.0 & 45.8 & 2.8          \\
  F02054+0835 & 0.58: & 2.42: & $-$0.06 & 0.63:& 0.28: & $<$0.16 & 0.04: & 169.3: & 33.4: &  80.9: &  9.2: & $<$3.1    \\
  F02065+4705 & 0.25 & 1.65 &    0.12 &   0.83& 0.35 & $<$0.06 & 0.01 & 189.7 & 38.4 &  63.5 & 13.6 & $<$1.4      \\
  F04415+1215 & 0.11 & 1.74 & 0.0 &   0.72& 0.55 & $<$0.05 & 0.12 & 209.7 & 50.9 &  88.3 & 28.0 & $<$1.6    \\
 IR06269$-$0542 & 0.39 & 0.57 &    0.0 &   1.54& 1.02 & 0.21 & $<$0.01 & 404.0 & 83.6 &  47.6 & 85.2 &12.2            \\
  F07599+6508 & 0.00 & 2.75 & $-$0.21 &$<$0.08& $<$0.02 & $<$0.04 & $<$0.01 & 258.3 & 47.2 & 129.6 & $<$1.1 & $<$1.3 \\
  F09427+1929 & 0.00 & 2.98 & $-$0.09 &$<$0.08& $<$0.02 & $<$0.02 & 0.08 & 222.0 & 52.0 & 155.1 & $<$ 1.1 & $<$0.6 \\
  F10026+4347 & 0.00 & 2.07 &    0.14 &   0.19& 0.10 & $<$0.01 & $<$0.01 & 149.4 & 36.2 &  75.0 &  3.8 & $<$0.1      \\
  F11119+3257 & 1.08 & 1.12 &    0.0 &   1.01& 0.75 & 0.64 & 0.12 & 181.8 & 51.1 &  57.2 & 38.5 &25.9            \\
Z11598$-$0112 & 0.01 & 1.71 &    0.81 &   1.10& 0.75 & 0.88 & 0.26 &  69.8 & 13.9 &  24.0 & 10.5 & 8.6            \\
  F12134+5459 & 0.14 & 1.28 &    0.0 &   1.13& 0.68 & 0.12 & 0.17 & 148.4 & 33.4 &  42.7 & 22.8 & 7.6            \\
  F12265+0219 & 0.00 & 0.66 &    0.0 &   0.21& 0.06 & $<$0.02 & 0.07 & 299.8 & 78.3 &  51.4 &  4.9 & $<$0.8      \\
  F12540+5708 & 0.65 & 1.83 & $-$0.25 &$<$0.07& $<$0.01 & $<$0.13 & 0.04 & 257.2 & 35.0 &  60.0 &$<$4.1 & $<$3.2  \\
  F13218+0552 & 0.54 & 0.48\tablenotemark{a} &...&   1.62:& 2.40\tablenotemark{a} & ...  & 0.28: & 162.7 & ...  &  ...  & 47.7: & ... \\
  F13342+3932 & 0.76 & 0.73 &    0.0 &   3.67& 1.09 & 0.83 & 0.11 & 203.2 & 35.0 &  25.6 & 38.0 &17.5            \\
  F15069+1808 & 0.01 & 0.92 &    0.0 &   2.04& 1.03 & 0.44 & 0.18 & 189.0 & 46.0 &  42.2 & 47.2 &13.7           \\
F15462$-$0450 & 0.28 & 1.32 &    0.0 &   0.73& 0.66 & 0.48 & 0.30 & 126.7 & 48.7 &  64.3 & 32.2 &13.5           \\
  F16136+6550 & 0.17 & 0.43 & 0.0 &   1.52& 0.20 & $<$0.03 & $<$0.01 & 325.3 & 60.0 &  26.0 & 11.7 & $<$1.1   \\
  F18216+6419 & 0.00 & 0.39 & 0.0 &   1.63& 0.29 & 0.04 & 0.04 & 435.6 & 79.9 &  30.9 & 22.9 & 1.7         \\
F20036$-$1547 & 0.20 & 2.74 & $-$0.09 &$<$0.10& $<$0.01 & $<$0.15 & 0.19 & 122.9 & 24.2 &  66.4 &$<$0.3 & $<$2.4 \\
F20520$-$2329 & 0.46 & 2.02 &    0.10 &   0.72& 0.37 & 0.12 & 0.33 & 103.2 & 28.0 &  56.5 & 10.3 & 2.3           \\
F21219$-$1757 & 0.23 & 1.82 &    0.05 &   0.49& 0.25 & $<$0.07 & 0.32 & 153.8 & 44.2 &  80.5 & 11.2 & $<$1.7     \\
F22454$-$1744 & 0.21 & 1.03 &    0.36 &   1.52& 1.08 & 0.71 & 0.17 & 166.4 & 33.2 &  34.3 & 35.8 &15.3           \\
  F23411+0228 & 0.71: & ...  &     ... &   ... & ...  & ...  & 0.08 &  59.0 &  ... &  ... & ... &  ...  \\

\noalign{\smallskip}
\tableline
\noalign{\smallskip}
\end{tabular}
\tablenotetext{a}{from Remillard et al. 1993.}
\tablecomments{
The typical uncertainty of the emission line fluxes is about 10--20\%. 
The values flagged with colon have uncertainties larger than
40\%. Upper limits are given if the emission lines are not convincingly
detected. No measurement is listed for the emission lines 
contaminated by noise.
Col.(1): source name, same as in table 1;
Col.(2): color excess;
Col.(3): strength of \Few\ relative to \Hb;
Col.(4): relative strength of \feii\ residual, positive
indicates excess of \feii\,37,38 and negative indicates excess of
\feii\,48,49 (cf Fig.\ref{FeIIresiduals});
Col.(5): peak height of \OIII\  relative to that of \Hb;
Col.(6): strength of \OIII\ relative to \Hb;
Col.(7): strength of \OII\ relative to \Hb;
Col.(8): strength of \NII$\lambda$6583 relative to \Ha;
Col.(9): equivalent width of \Ha;
Col.(10): equivalent width of \Hb;
Col.(11): equivalent width of \Few;
Col.(12): equivalent width of \OIII;
Col.(13): equivalent width of \OII.
}
\end{center}
\end{table}

\end{document}